\documentclass[12pt,preprint]{aastex}

\slugcomment{to appear in the \emph{Astrophysical Journal}}

\shorttitle{Sulfur and Phosphorus in Substellar Objects}

\shortauthors{Visscher et al.}

\begin{document}

\title{Atmospheric Chemistry in Giant Planets, Brown Dwarfs, and Low-Mass Dwarf Stars II. Sulfur and Phosphorus}

\author{Channon Visscher, Katharina Lodders, and Bruce Fegley, Jr.}
\affil{Planetary Chemistry Laboratory, Department of Earth \& Planetary Sciences, McDonnell Center for the Space Sciences, Washington University,
St. Louis, MO 63130-4899}

\email{visscher@wustl.edu, lodders@wustl.edu, bfegley@wustl.edu}

\begin{abstract}
Thermochemical equilibrium and kinetic calculations are used to model sulfur and phosphorus chemistry in giant planets, brown dwarfs, and
extrasolar giant planets (EGPs).  The chemical behavior of individual S- and P-bearing gases and condensates is determined as a function of
pressure, temperature, and metallicity.  The results are independent of particular model atmospheres and, in principle, the equilibrium
composition along the pressure-temperature profile of any object can be determined.  Hydrogen sulfide (H$_{2}$S) is the dominant S-bearing gas
throughout substellar atmospheres and approximately represents the atmospheric sulfur inventory.  Silicon sulfide (SiS) is a potential tracer of
weather in substellar atmospheres. Disequilibrium abundances of phosphine (PH$_{3}$) approximately representative of the total atmospheric
phosphorus inventory are expected to be mixed upward into the observable atmospheres of giant planets and T dwarfs.  In hotter objects, several
P-bearing gases (e.g., P$_{2}$, PH$_{3}$, PH$_{2}$, PH, HCP) become increasingly important at high temperatures.
\end{abstract}


\section{Introduction}\label{s Introduction}

The relatively cool, dense atmospheres of substellar objects -- gas giant planets, brown dwarfs, and extrasolar giant planets (EGPs) -- are ideal
environments for the formation of molecules and condensates and the establishment of equilibrium chemistry.  For this reason, thermochemical
models have been essential for interpreting and guiding spectroscopic observations of the atmospheres of giant planets \citep*[e.g., see][]{lewis
1969a, lewis 1969b, barshay and lewis 1978, fegley and lewis 1979, fegley and prinn 1985, fegley and prinn 1988, fegley et al 1991, fegley and
lodders 1994} and brown dwarfs \citep*[e.g.,][]{fegley and lodders 1996}. These models have also been useful for explaining how the optical and
infrared spectra of substellar objects are influenced by CNO \citep*{lodders and fegley 2002}, alkali \citep{lodders 1999a}, Ti and V
\citep{lodders 2002}, and condensation \citep[e.g.,][]{lodders and fegley 2006} chemistry.

After H, C, N, and O, sulfur and phosphorus are the next most abundant chemically reactive volatile elements in a solar system composition gas.
Fegley \& Lodders (1994; hereafter FL94) modeled the chemistry of all naturally occurring elements, including S and P, along the atmospheric
profiles of Jupiter and Saturn. Recently, Lodders \& Fegley (2002; hereafter LF02) modeled CNO chemistry in substellar objects as a function of
pressure, temperature, and metallicity.  Here we continue and extend these studies by employing thermochemical equilibrium and kinetic
calculations to model sulfur and phosphorus chemistry in the atmospheres of substellar objects.

Our approach is similar to that of \citet{lodders and fegley 2002}.  We compute the chemistry of S- and P-bearing gases and condensates as a
function of total pressure, temperature, and metallicity, and our abundance results are \textit{independent} of any particular atmospheric
profile. In principle, the pressure-temperature profile for any substellar object may be superimposed on our abundance contour diagrams to find
the equilibrium composition along the profile.

The paper is organized as follows.  In section ($\S$\ref{s Computational Method}) we describe our computational methods.  In $\S$\ref{s Sulfur
Chemistry} we describe sulfur chemistry in a solar system composition gas: we identify major gases and condensates, and discuss the chemistry of
individual S-bearing species as a function of total pressure, temperature, and metallicity. We then apply our results by examining the sulfur
species along the atmospheric profiles of representative substellar objects. A similar treatment of phosphorus chemistry follows in $\S$\ref{s
Phosphorus Chemistry}.  We discuss the relative roles of thermochemistry and photochemistry in $\S$\ref{s Photochemistry}, and conclude with a
brief summary ($\S$\ref{s Summary}).

\section{Computational Methods}\label{s Computational Method}

Thermochemical equilibrium calculations were performed using a Gibbs free energy minimization code, previously used for modeling the atmospheric
chemistry of Saturn \citep{visscher and fegley 2005}.  Where relevant, we considered the effects of vertical mixing on the abundances of gases
(e.g., PH$_{3}$) which serve as chemical probes of the deep atmospheres of Jupiter and Saturn.  This was done using a chemical dynamical model
described in \citet{fegley and prinn 1985}.  We generally focus on temperatures of 1000 K and higher, where thermochemistry is expected to be much
more important than photochemistry in the atmospheres of extrasolar giant planets close to their primary stars. Further discussion of
thermochemical vs.~photochemical processes is given in \S\ref{s Photochemistry}.

Thermodynamic data for the equilibrium calculations were taken from the compilations of \citet{gurvich et al 1989-1994}, \citet{robie and
hemingway 1995}, the fourth edition of the JANAF Tables \citep{chase 1999}, and the thermodynamic database maintained in the Planetary Chemistry
Laboratory \citep{fegley and lodders 1994, lodders and fegley 2002}. This database includes additional thermodynamic data from the literature for
compounds absent from the other compilations, as well as several important data revisions for the sulfur and phosphorus species SH, S$_{2}$O, NS,
PS, PH, PH$_{3}$, PN, and Mg$_{3}$P$_{2}$O$_{8}$ (s), which are incorrect in the JANAF tables \citep{lodders 1999b, lodders 2004a}. Thermodynamic
data for P$_{4}$O$_{6}$ was taken from the JANAF tables for reasons cited in \citet{fegley and lodders 1994}.

All calculations were done with solar system (i.e., protosolar) elemental abundances from \citet{lodders 2003}. The effect of metallicity on
sulfur and phosphorus chemistry was examined by running computations at [M/H] = -0.5 dex (subsolar), [M/H] = 0 dex (solar), and [M/H] = +0.5 dex
(enhanced) metallicities, where M is any element of interest (e.g., S, P, C, O)\footnote{We use the conventional notation [M/H] $\equiv \log
(\textrm{M/H}) - \log (\textrm{M/H})_{\textrm{\scriptsize{solar}}}$.}. The metallicity factor, $m$, is defined as $\log m=[\textrm{M/H}]$. Stellar
abundance determinations show that element abundance ratios [M/H] for elements of interest here vary similarly with [Fe/H] between $-0.5 \leq
[\textrm{Fe/H}] \leq +0.5$ dex \citep{edvardsson et al 1993, gustafsson et al 1999, chen et al 2000, smith et al 2001, ecuvillon et al 2004, huang
et al 2005}, consistent with what one may expect from galactic chemical evolution models \citep[e.g.,][]{timmes et al 1995}.  The slopes for fits
of [M/H] vs.~[Fe/H] are about 0.61, 0.65, and 0.6 for M = C, S, and O, respectively, which justifies our approach of ``uniform" metallicity
variations for these elements (e.g., [S/H] $\approx$ [P/H] $\approx$ [C/H] $\approx$ [O/H] $\equiv$ [M/H]) over the metallicity range considered
here.  Because phosphorus has weak lines, galactic chemical evolution models are used to estimate P abundances as a function of metallicity
\citep[e.g.,][]{timmes et al 1995, samland 1998}. We thus assume a similar enrichment in phosphorus as for other heavy elements. However, the
effects of metallicity on chemical equilibrium abundances are explicitly treated below and can be used to study effects from non-uniform elemental
abundance enrichments.

For each S- and P-bearing species, we discuss how the most plausible representative net thermochemical formation reaction depends on temperature,
total pressure, and metallicity.  For example, consider the formation of thioxophosphino (PS) via the net thermochemical reaction
\begin{displaymath}
\textrm{PH}_{3}+\textrm{H}_{2}\textrm{S}=\textrm{PS}+2.5\textrm{H}_{2}
\end{displaymath}
This endothermic reaction proceeds toward the right with increasing temperature and the PS abundance increases.  According to LeCh\^{a}telier's
principle, this reaction also proceeds toward the right with decreasing pressure because there are 2 gas molecules on the left and 3.5 gas
molecules on the right. Writing out the equilibrium constant ($K_{p}$) expression for PS in terms of the total pressure ($P_{T}$) and mole
fractions\footnote{The mole fraction abundance for a gas $i$ is defined as $X_{i}= \frac{\textrm{\scriptsize{moles }}i}{\textrm{\scriptsize{total
moles of all gases}}}$.}
\begin{displaymath}
X_{\textrm{\scriptsize{PS}}}=(X_{\textrm{\scriptsize{PH}}_{3}}X_{\textrm{\scriptsize{H}}_{2}\textrm{\scriptsize{S}}}/X_{\textrm{\scriptsize{H}}_{2}}^{2.5})K_{p}
P_{T}^{-1.5},
\end{displaymath}
shows that the mole fraction of PS is proportional to $P_{T}^{-1.5}$.  The metallicity dependence of $X_{\textrm{\scriptsize{PS}}}$ enters only
through the PH$_{3}$ and H$_{2}$S abundances, whereas the H$_{2}$ abundance is metallicity-independent by definition.  Phosphine and H$_{2}$S are
typically the most abundant P- and S-bearing gases, respectively. Hence $X_{\textrm{\scriptsize{PH}}_{3}}$ and
$X_{\textrm{\scriptsize{H}}_{2}\textrm{\scriptsize{S}}}$ must each be expanded with a metallicity factor $m$, which results in an overall $m^{2}$
metallicity dependence for $X_{\textrm{\scriptsize{PS}}}$, assuming [S/H] = [P/H] (see above). This basic approach is used to describe the
equilibrium chemical behavior of each S- and P-bearing species throughout the paper.

\section{Sulfur Chemistry}\label{s Sulfur Chemistry}
\subsection{Sulfur Gas Chemistry}\label{ss Overview of Sulfur Chemistry}
\begin{figure}
\scalebox{0.46}{\includegraphics{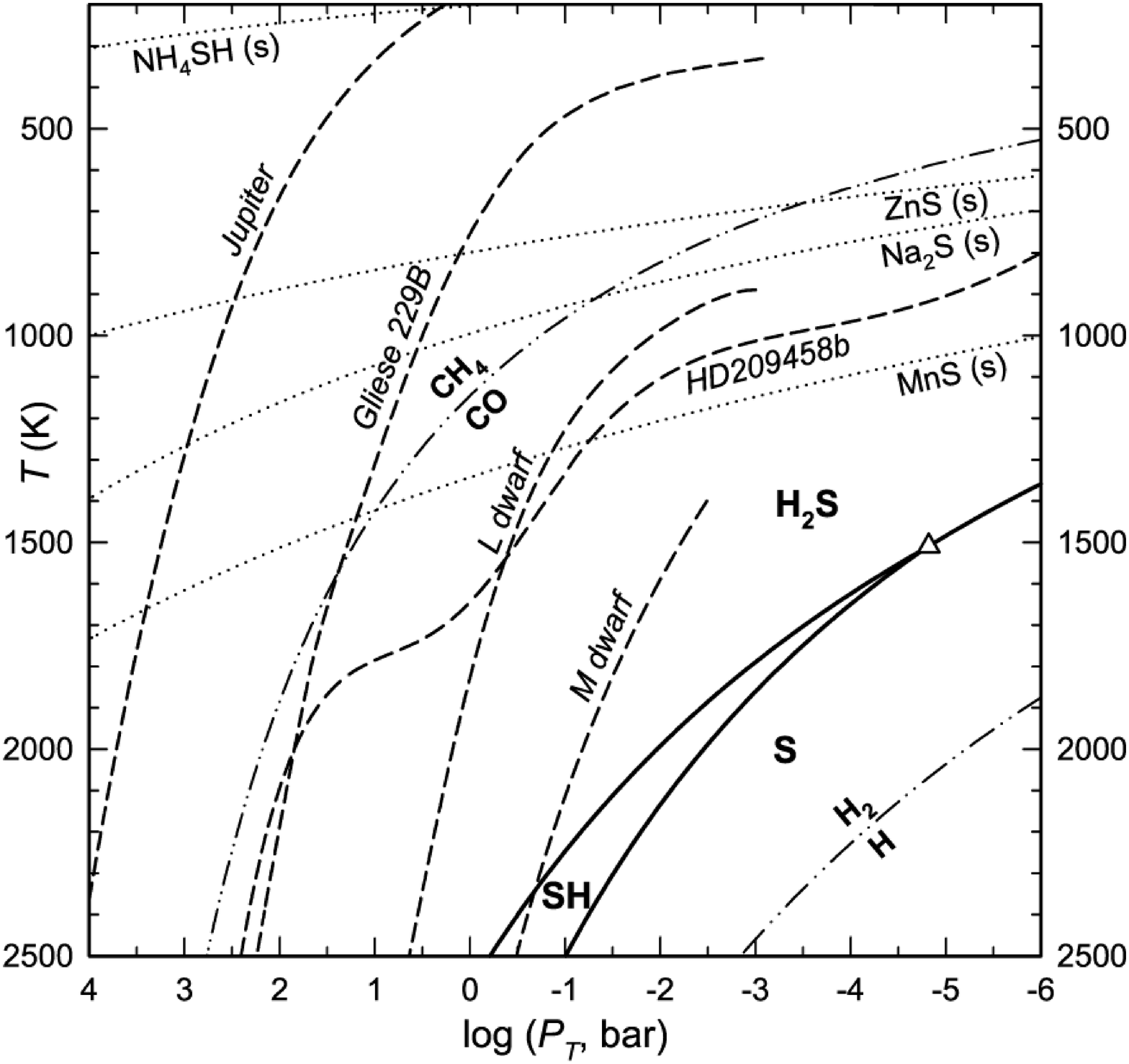}} \caption[Overview of Sulfur Chemistry]{Overview of sulfur chemistry as a
function of temperature and pressure in a solar system composition gas.  The solid lines indicate where major
S-bearing gases have equal abundances.  The triangle denotes the position of the triple point where H$_{2}$S, SH,
and S have equal abundances.  Also shown are condensation curves for S-bearing compounds (dotted lines) and the
H$_{2}$=H and CH$_{4}$=CO equal abundance lines (dash-dot lines).  Atmospheric profiles (dashed lines) are given
for Jupiter ($T_{\textrm{\scriptsize{eff}}}$ = 124 K), Gliese 229B ($T_{\textrm{\scriptsize{eff}}}$ = 960 K),
HD209458b ($T_{\textrm{\scriptsize{eff}}}$ = 1350 K), an L dwarf ($T_{\textrm{\scriptsize{eff}}}$ = 1800 K), and
an M dwarf ($T_{\textrm{\scriptsize{eff}}}$ = 2200 K).} \label{figure sulfur chemistry}
\end{figure}

The sulfur equilibrium gas chemistry as a function of temperature and total pressure in a protosolar composition gas is illustrated in Figure
\ref{figure sulfur chemistry}. The most abundant gases are H$_{2}$S, SH, or monatomic S, and the \textit{P-T} regions where each gas is dominant
are bounded by solid lines. Also shown are condensation curves (dotted lines) for S-bearing condensates (see $\S$\ref{ss Sulfur Condensation
Chemistry}) and the H$_{2}$=H and CH$_{4}$=CO equal abundance curves (dash-dot lines). Model atmosphere profiles for Jupiter
($T_{\textrm{\scriptsize{eff}}}$ = 124 K, $\log g=3.4$), the T dwarf Gliese 229B \citep[$T_{\textrm{\scriptsize{eff}}}$ = 960 K, $\log
g=5.0$;][]{marley et al 1996}, the close-orbiting EGP (or ``Pegasi" planet) HD209458b \citep[$T_{\textrm{\scriptsize{eff}}}$ = 1350 K, $\log
g=3.0$;][]{iro et al 2005}, an L dwarf \citep[$T_{\textrm{\scriptsize{eff}}}$ = 1800 K, $\log g=5.0$;][]{burrows et al 2006}, and an M dwarf
\citep[$T_{\textrm{\scriptsize{eff}}}$ = 2600 K, dust-free, $\log g=5.0$;][]{tsuji et al 1996} are indicated by dashed lines. The overall
chemistry in the atmospheres of Jupiter and Gliese 229B differs slightly from the chemistry of a solar metallicity gas because Jupiter has a heavy
element enrichment comparable to [M/H] $\approx$ +0.5 dex \citep{lodders 1999a, lodders and fegley 2002}, whereas Gliese 229B likely has a
subsolar metallicity of [M/H] $\approx$ -0.3 dex \citep{saumon et al 2000}; their \textit{P-T} profiles are shown here for reference. However, as
described below, the \textit{P-T} boundaries for the sulfur gases in Figure \ref{figure sulfur chemistry} are metallicity independent. This is
because the \textit{P-T} boundaries are defined by equal abundance \textit{ratios} of the neighboring gases (e.g., H$_{2}$S/SH, SH/S, H$_{2}$S/S),
so that the metallicity dependence cancels out.

Hydrogen sulfide (H$_{2}$S) is the dominant S-bearing gas in substellar atmospheres under most \textit{P-T} conditions considered here. The
absolute H$_{2}$S abundance is of course metallicity dependent (see $\S$\ref{sss Hydrogen Sulfide, H2S}).  At high temperatures and low pressures,
H$_{2}$S is replaced by SH via the net thermochemical reaction
\begin{equation}\label{equationH2S:SH}
\textrm{H}_{2}\textrm{S}=\textrm{SH}+0.5\textrm{H}_{2},
\end{equation}
and SH becomes the dominant sulfur gas.  The equilibrium constant expression for reaction (\ref{equationH2S:SH}) is
\begin{equation}
K_{\ref{equationH2S:SH}}=(X_{\textrm{\scriptsize{SH}}}X_{\textrm{\scriptsize{H}}_{2}}^{0.5}/X_{\textrm{\scriptsize{H}}_{2}\textrm{\scriptsize{S}}})P_{T}^{0.5}.
\end{equation}
Rearranging and substituting for the temperature dependence of $K_{\ref{equationH2S:SH}}$ ($\log K_{\ref{equationH2S:SH}} = 3.37-8785/T$ from 800
to 2500 K), the H$_{2}$S/SH ratio is given by
\begin{equation}\label{ratio H2S:SH}
\log(X_{\textrm{\scriptsize{H}}_{2}\textrm{\scriptsize{S}}}/X_{\textrm{\scriptsize{SH}}})=-3.37+8785/T+0.5\log P_{T}+0.5 \log
X_{\textrm{\scriptsize{H}}_{2}}.
\end{equation}
The H$_{2}$S/SH ratio is independent of metallicity because the $m$ dependence of each S-bearing gas in reaction (\ref{equationH2S:SH}) cancels
out and the H$_{2}$ abundance is essentially constant ($X_{\textrm{\scriptsize{H}}_{2}}\approx0.84$) over small metallicity variations.  The solid
line separating the H$_{2}$S and SH fields in Figure \ref{figure sulfur chemistry} shows where these two gases have equal abundances. With
$X_{\textrm{\scriptsize{H}}_{2}\textrm{\scriptsize{S}}}/X_{\textrm{\scriptsize{SH}}}=1$ and $X_{\textrm{\scriptsize{H}}_{2}}\approx0.84$, equation
(\ref{ratio H2S:SH}) can be rewritten to give the H$_{2}$S=SH equal abundance boundary:
\begin{equation}\label{equation line H2S:SH}
\log P_{T}=6.82-17570/T,
\end{equation}
which is independent of metallicity.  The abundance of each sulfur gas does not drop to zero as this line is crossed; H$_{2}$S is still present
within the SH field and vice versa (see $\S$\ref{sss Mercapto, SH}).

With increasing temperatures, SH dissociates and monatomic S becomes the dominant sulfur-bearing gas via
\begin{equation}\label{equationSH:S}
\textrm{SH}=\textrm{S}+0.5\textrm{H}_{2}.
\end{equation}
Using $\log K_{\ref{equationSH:S}}=2.36-7261/T$ from 800 to 2500 K and rearranging, the SH/S abundance ratio is given by
\begin{equation}\label{ratio SH:S}
\log(X_{\textrm{\scriptsize{SH}}}/X_{\textrm{\scriptsize{S}}})=-2.36+7261/T+0.5\log P_{T}+0.5 \log X_{\textrm{\scriptsize{H}}_{2}}.
\end{equation}
This ratio is also independent of metallicity because the $m$ dependence of each S-bearing gas in reaction (\ref{equationSH:S}) cancels out. With
$X_{\textrm{\scriptsize{SH}}}/X_{\textrm{\scriptsize{S}}}=1$ and $X_{\textrm{\scriptsize{H}}_{2}}\approx 0.84$, equation (\ref{ratio SH:S}) is
rewritten to give the position of the metallicity-independent SH=S equal abundance line
\begin{equation}\label{equation line SH:S}
\log P_{T}=4.80-14522/T.
\end{equation}

At the lowest total pressures considered here, monatomic S is the major sulfur gas (see Figure \ref{figure sulfur chemistry}).  At temperatures
below $\sim1509$ K, H$_{2}$S directly converts to monatomic S via the net thermochemical reaction
\begin{equation}\label{equationH2S:S}
\textrm{H}_{2}\textrm{S}=\textrm{S}+\textrm{H}_{2}.
\end{equation}
The H$_{2}$S/S abundance ratio follows from adding equations (\ref{ratio H2S:SH}) and (\ref{ratio SH:S}) as
\begin{equation}\label{ratio H2S:S}
\log(X_{\textrm{\scriptsize{H}}_{2}\textrm{\scriptsize{S}}}/X_{\textrm{\scriptsize{S}}})=-5.73+16046/T+\log P_{T}+\log
X_{\textrm{\scriptsize{H}}_{2}},
\end{equation}
and is independent of metallicity. Again, with $X_{\textrm{\scriptsize{H}}_{2}\textrm{\scriptsize{S}}}/X_{\textrm{\scriptsize{S}}}=1$ and
$X_{\textrm{\scriptsize{H}}_{2}}\approx 0.84$, the metallicity-independent position of the H$_{2}$S=S boundary is approximated by
\begin{equation}\label{equation line H2S:S}
\log P_{T}=5.81-16046/T.
\end{equation}
The H$_{2}$S=SH, SH=S, and H$_{2}$S=S boundaries intersect at the H$_{2}$S-SH-S ``triple point'' at $T \sim 1509$ K and $P_{T} \sim 10^{-4.82}$
bar, indicated by the triangle in Figure \ref{figure sulfur chemistry} and represented by the net thermochemical equilibrium
\begin{equation}\label{equationH2S:SH:S}
\textrm{H}_{2}\textrm{S}+\textrm{S}=2\textrm{SH}.
\end{equation}
The \textit{P-T} position of the triple point is independent of metallicity because the intersecting boundaries in equations (\ref{equation line
H2S:SH}), (\ref{equation line SH:S}), and (\ref{equation line H2S:S}) are each independent of metallicity. At the triple point, all three gases
have equal abundances A(H$_{2}$S) = A(SH) = A(S) $\approx$ $\frac{1}{3}\Sigma\textrm{S}$ (where $\Sigma$S is the total elemental sulfur abundance)
of one third of the total sulfur abundance.

As noted above, the H$_{2}$ abundance is essentially constant ($X_{\textrm{\scriptsize{H}}_{2}}\approx 0.84$) over the small metallicity range
considered here.  However, at high $T$ and low $P_{T}$, H$_{2}$ thermally dissociates to H via the net thermochemical reaction
\begin{equation}\label{equation H2:H}
\textrm{H}_{2}=2\textrm{H}.
\end{equation}
The position of the H$_{2}$=H equal abundance boundary is indicated by the dash-dot line in Figure \ref{figure sulfur chemistry}.  The abundance
of H$_{2}$ as a function of $P_{T}$ and $T$ in a solar system composition gas can be determined by solving the expression
\begin{equation}\label{equation H2 quadratic}
X_{\textrm{\scriptsize{H}}_{2}} \approx \frac{1.9845 + 10^{u} - \sqrt{10^{u}(3.9690+10^{u})}}{2.3670},
\end{equation}
where the variable $u$ is given by
\begin{equation}\label{equation H2 PT dependence}
u=-23672/T-\log P_{T}+6.2645.
\end{equation}
At pressures and temperatures relevant for substellar atmospheres (Figure \ref{figure sulfur chemistry}), $10^{u}$ approaches zero and
$X_{\textrm{\scriptsize{H}}_{2}}\approx 0.84$.

\subsubsection{Hydrogen Sulfide, H$_{\textrm{\scriptsize{2}}}$S gas}\label{sss Hydrogen Sulfide, H2S}
\begin{figure}
\scalebox{.43}{\includegraphics{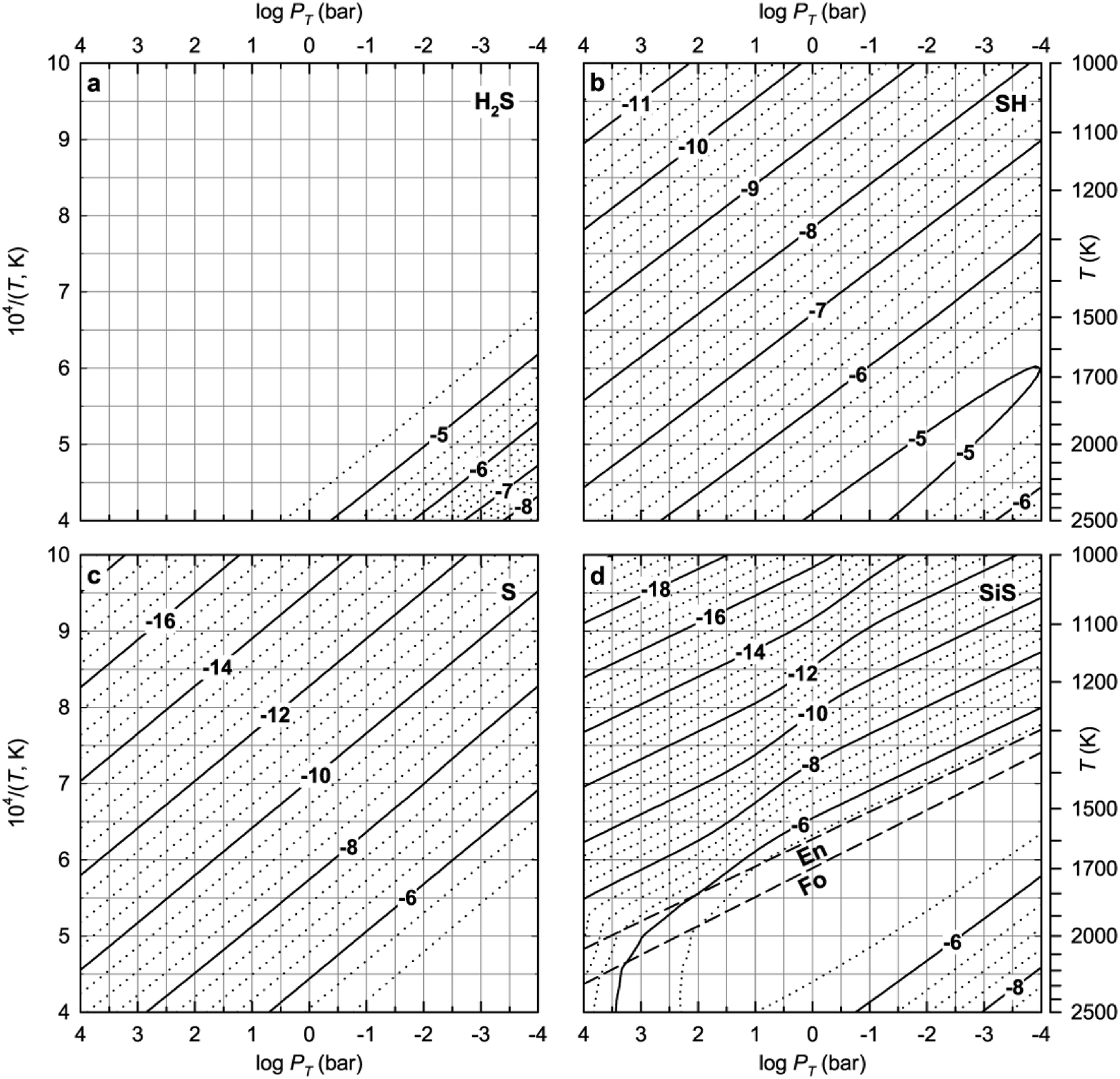}} \caption[Individual sulfur gases]{Mole fraction contours (on a
logarithmic scale) for (a) hydrogen sulfide (H$_{2}$S), (b) mercapto radical (SH), (c) monatomic sulfur (S), (d)
silicon sulfide (SiS) as a function of pressure and temperature in a solar metallicity gas.  The dashed lines
labeled Fo and En in the SiS plot indicate the equilibrium condensation temperatures of forsterite
(Mg$_{2}$SiO$_{4}$) and enstatite (MgSiO$_{3}$), respectively.} \label{figure individual S gas}
\end{figure}

As shown in Figure \ref{figure sulfur chemistry}, H$_{2}$S is expected to be the most abundant S-bearing gas throughout the atmospheres of
substellar objects \citep[e.g.][]{fegley and lodders 1994, fegley and lodders 1996}.  Aside from minor amounts of sulfur ($\lesssim 9\%$) removed
by metal sulfide clouds (see $\S$\ref{ss Sulfur Condensation Chemistry}), H$_{2}$S approximately represents the atmospheric sulfur inventory until
it is removed by very low temperature cloud formation and/or photochemical destruction.  The H$_{2}$S abundance also slightly decreases at
atmospheric levels below the Mg-silicate clouds where SiS gas is relatively abundant (see $\S$\ref{sss Silicon Sulfide, SiS} and $\S$\ref{ss
Sulfur Chemistry in Substellar Objects}).  Within the H$_{2}$S field and at the high temperatures below the metal sulfide clouds, the H$_{2}$S
abundance is given by
\begin{equation}\label{equation H2S in H2S}
\log X_{\textrm{\scriptsize{H}}_{2}\textrm{\scriptsize{S}}}\approx -4.52+[\textrm{S/H}],
\end{equation}
whereas above the metal sulfide clouds, the H$_{2}$S abundance is
\begin{equation}\label{equation H2S in H2S above clouds}
\log X_{\textrm{\scriptsize{H}}_{2}\textrm{\scriptsize{S}}}\approx -4.56+[\textrm{S/H}].
\end{equation}
Figure \ref{figure individual S gas}a gives mole fraction contours (on a logarithmic scale) of H$_{2}$S as a function of temperature and total
pressure in a solar metallicity gas. The H$_{2}$S abundance at higher or lower metallicities can be found by substituting for [S/H] in equations
(\ref{equation H2S in H2S}) and (\ref{equation H2S in H2S above clouds}). As discussed above, H$_{2}$S dissociates to SH and S at high
temperatures. Inside the SH and S fields, the H$_{2}$S abundance remains proportional to $m$ and decreases when moving toward higher temperatures
and lower total pressures.

\subsubsection{Mercapto, SH gas}\label{sss Mercapto, SH}

Mole fraction contours of SH as a function of $T$ and $P_{T}$ are shown in Figure \ref{figure individual S gas}b. Within the H$_{2}$S field
($X_{\textrm{\scriptsize{H}}_{2}\textrm{\scriptsize{S}}}\approx X_{\Sigma\textrm{\scriptsize{S}}}$ and $X_{\textrm{\scriptsize{H}}_{2}}\approx
0.84$), mercapto is governed by reaction (\ref{equationH2S:SH}).  The SH abundance is given by substituting equation (\ref{equation H2S in H2S})
into equation (\ref{ratio H2S:SH}) and rearranging
\begin{equation}\label{equation SH in H2S}
\log X_{\textrm{\scriptsize{SH}}} \approx -1.11-8785/T-0.5\log P_{T}+[\textrm{S/H}],
\end{equation}
proportional to $P_{T}^{-0.5}$ and $m$.  At high temperatures, SH is replaced by S via reaction (\ref{equationSH:S}).  Within the S field, the SH
abundance is proportional to $m$ and decreases when moving away from the SH-S boundary to higher $T$ and lower $P_{T}$.

\subsubsection{Monatomic Sulfur, S gas}\label{sss Monatomic Sulfur, S}

Mole fraction contours for monatomic S are shown in Figure \ref{figure individual S gas}c.  Inside the H$_{2}$S field, monatomic sulfur gas is
governed by reaction (\ref{equationH2S:S}).  The S abundance is given by rearranging equation (\ref{ratio H2S:S}), using $\log
X_{\textrm{\scriptsize{H}}_{2}\textrm{\scriptsize{S}}}$ from equation (\ref{equation H2S in H2S}) and $\log
X_{\textrm{\scriptsize{H}}_{2}}\approx-0.08$ to give
\begin{equation}\label{equation S in H2S}
\log X_{\textrm{\scriptsize{S}}}\approx 1.29-16046/T-\log P_{T}+[\textrm{S/H}],
\end{equation}
proportional to $P_{T}^{-1}$ and $m$.  As illustrated in Figure \ref{figure sulfur chemistry}, monatomic S is the dominant S-bearing gas
($X_{\textrm{\scriptsize{S}}}\approx X_{\Sigma\textrm{\scriptsize{S}}}$) at high $T$ and low $P_{T}$ in a solar composition gas. Thermal
ionization of S only becomes important at temperatures higher than those shown in Figure \ref{figure sulfur chemistry}, e.g.,
$\textrm{S}^{+}/\textrm{S}\sim1$ at 6974 K ($10^{-2}$ bars), 5340 K ($10^{-4}$ bars), and 4326 K ($10^{-6}$ bars).

\subsubsection{Silicon Sulfide, SiS gas}\label{sss Silicon Sulfide, SiS}
The chemistry of SiS in a solar metallicity gas is shown in Figure \ref{figure individual S gas}d.  Also shown are dashed lines for the
condensation curves of forsterite (Mg$_{2}$SiO$_{4}$) and enstatite (MgSiO$_{3}$).  Condensation of Mg-silicates efficiently removes Si gases from
the atmospheres of gas giant planets and cool brown dwarfs \citep{fegley and prinn 1988, fegley and lodders 1994}. The SiS abundance is controlled
by the net thermochemical reaction
\begin{equation}\label{reaction SiO+H2S=SiS+H2O}
\textrm{SiO}+\textrm{H}_{2}\textrm{S}=\textrm{SiS}+\textrm{H}_{2}\textrm{O}.
\end{equation}
The SiS abundance at high temperatures below the Mg-silicate clouds is given by
\begin{equation}\label{equation SiS below clouds}
\log X_{\textrm{\scriptsize{SiS}}}\approx -1.07+666/T+[\textrm{M/H}]+\log X_{\textrm{\scriptsize{H}}_{2}\textrm{\scriptsize{S}}},
\end{equation}
assuming [O/H] = [S/H] (see $\S$\ref{s Computational Method}).  Below the clouds, $X_{\textrm{\scriptsize{SiS}}}$ is relatively constant ($\sim$ 5
ppm at [M/H] = 0) until H$_{2}$S is replaced by monatomic S at high $T$ and low $P_{T}$.  The SiS abundance also decreases at $P_{T}\gtrsim100$
bar as CO is replaced by CH$_{4}$ (see Figure \ref{figure sulfur chemistry}).  This conversion increases the H$_{2}$O abundance
\citep[see][]{lodders and fegley 2002}, and drives reaction (\ref{reaction SiO+H2S=SiS+H2O}) toward the left to yield less SiS.

At lower temperatures, above the clouds, the SiO abundance is governed by the reaction
\begin{equation}\label{reaction Mg2SiO4+SiO+H2O=2MgSiO3+H2}
\textrm{Mg}_{2}\textrm{SiO}_{4}(\textrm{s})+\textrm{SiO}+\textrm{H}_{2}\textrm{O}=2\textrm{MgSiO}_{3}(\textrm{s})+\textrm{H}_{2}.
\end{equation}
Combined with reaction (\ref{reaction SiO+H2S=SiS+H2O}), the abundance of SiS is given by
\begin{equation}\label{K express SiS abundance CH4}
\log X_{\textrm{\scriptsize{SiS}}}\approx 5.38-28151/T-\log P_{T}-2\log X_{\textrm{\scriptsize{H}}_{2}\textrm{\scriptsize{O}}}+[\textrm{S/H}].
\end{equation}
Although reaction (\ref{reaction SiO+H2S=SiS+H2O}) does not indicate a dependence on $P_{T}$ and suggests a linear dependence on the H$_{2}$O
abundance, the SiS abundance in equation (\ref{K express SiS abundance CH4}) is proportional to $P_{T}^{-1}$ and
$X_{\textrm{\scriptsize{H}}_{2}\textrm{\scriptsize{O}}}^{-2}$. These dependencies enter through the SiO abundance governed by reaction
(\ref{reaction Mg2SiO4+SiO+H2O=2MgSiO3+H2}). The H$_{2}$O abundance above the Mg-silicate clouds is approximated by
\begin{equation}\label{equation H2O abundance above clouds}
\log X_{\textrm{\scriptsize{H}}_{2}{\textrm{\scriptsize{O}}}}\approx-3.58+0.46/(1+10^{u})+[\textrm{O/H}],
\end{equation}
where
\begin{equation}
u=-11704/T-2\log P_{T}+9.78.
\end{equation}
Substituting $\log X_{\textrm{\scriptsize{H}}_{2}\textrm{\scriptsize{O}}}$ from equation (\ref{equation H2O abundance above clouds}) into equation
(\ref{K express SiS abundance CH4}) shows that $X_{\textrm{\scriptsize{SiS}}}$ is proportional to $m^{-1}$, assuming [O/H] = [S/H] (see $\S$\ref{s
Computational Method}). Curvature in the SiS abundance contours results from the dependence of the H$_{2}$O abundance on the CH$_{4}$/CO
equilibrium \citep[see][]{lodders and fegley 2002}. For reference, $X_{\textrm{\scriptsize{H}}_{2}\textrm{\scriptsize{O}}}\approx 10^{-3.12}$ in
CH$_{4}$-dominated objects and $X_{\textrm{\scriptsize{H}}_{2}\textrm{\scriptsize{O}}}\approx 10^{-3.58}$ in CO-dominated objects with solar
metallicity in atmospheric regions above the Mg-silicate clouds.

\subsection{Sulfur Condensation Chemistry}\label{ss Sulfur Condensation Chemistry}
With decreasing temperatures, sulfur condenses into MnS, Na$_{2}$S, ZnS, and NH$_{4}$SH cloud layers.  Condensation affects the spectra of
substellar objects by removing gases (and thus potential opacity sources) from the observable atmosphere, and by introducing cloud particles
\citep[][]{marley et al 1996, lodders 1999a, burrows et al 2000a, lodders and fegley 2006}.

The formation of a sulfur-bearing cloud requires that its constituent elements were not removed by condensation into more refractory cloud layers
at deeper levels.  For example, one might naively expect FeS cloud formation because FeS is a well-known condensate in the equilibrium
condensation sequence of the solar nebula gas. However, condensation of FeS is preceded by Fe metal condensation and in the high-gravity
environment of substellar objects, iron settles into a deep cloud layer at high temperatures \citep[e.g.][]{prinn and olaguer 1981, fegley and
lodders 1994, lodders 1999a, lodders and fegley 2002, lodders and fegley 2006}. Hence no Fe metal is left at lower temperatures when FeS could
condense.  Note that if FeS condensation did occur, all H$_{2}$S would be removed because Fe/S $\approx 1.9$. However, the measured H$_{2}$S
abundance of $\sim2.4$ times the protosolar H$_{2}$S/H$_{2}$ ratio in Jupiter's atmosphere \citep{niemann et al 1998, wong et al 2004} requires Fe
metal cloud formation at deep atmospheric levels.

Sulfur thus condenses as other sulfides with metals (Mn, Na, Zn) that are not already removed into high-temperature condensates.  The abundances
of Mn, Zn, and Na are all lower than that of S and condensation of these sulfides only depletes the atmosphere of $\sim9\%$ of its entire sulfur
inventory.  Therefore, H$_{2}$S gas remains in T dwarfs, L dwarfs, and Pegasi planets, and is only depleted in objects with low enough
temperatures for NH$_{4}$SH condensation.  Table \ref{table sulfur condensation} lists the calculated condensation temperatures of MnS, Na$_{2}$S,
ZnS, and NH$_{4}$SH for HD209458b and a model L dwarf with [M/H] $\approx$ 0, for Jupiter ([M/H] $\approx$ +0.5), and for Gliese 229B ([M/H]
$\approx$ -0.3), assuming uniform enrichments and/or depletions of heavy element abundances.

\subsubsection{Alabandite, MnS, condensation}\label{sss Manganese Sulfide, MnS}

The most refractory S-bearing condensate in substellar atmospheres is MnS (Figure \ref{figure sulfur chemistry}). Alabandite clouds form via the
net thermochemical reaction
\begin{equation}\label{equation MnS condensation reaction}
\textrm{Mn}+\textrm{H}_{2}\textrm{S}=\textrm{MnS(s)}+\textrm{H}_{2},
\end{equation}
where the condensation temperature is approximated by
\begin{eqnarray}\label{equation MnS condensation approximation}
10^{4}/T_{\textrm{\scriptsize{cond}}}(\textrm{MnS}) & \approx & 7.45-0.42(\log P_{T}+[\textrm{Mn/H}]+[\textrm{S/H}])\nonumber\\
& \approx & 7.45-0.42\log P_{T}-0.84[\textrm{M/H}],
\end{eqnarray}
assuming [Mn/H] = [S/H] (see $\S$\ref{s Computational Method}).  Because MnS condenses at sufficiently high temperatures (e.g. $\sim1340$ K at 1
bar), MnS clouds are expected in most substellar atmospheres.  The Mn abundance is 2\% that of sulfur and thus MnS cloud formation removes all Mn
and 2\% of all sulfur from the atmosphere.

\subsubsection{Sodium Sulfide, Na$_{\textrm{\scriptsize{2}}}$S, condensation}\label{sss Sodium Sulfide}

The Na$_{2}$S cloud layer forms via the net thermochemical reaction
\begin{equation}\label{equation Na2S condensation reaction}
2\textrm{Na}+\textrm{H}_{2}\textrm{S}=\textrm{Na}_{2}\textrm{S(s)}+\textrm{H}_{2},
\end{equation}
The condensation temperature of Na$_{2}$S as a function of $P_{T}$ and metallicity is approximated by
\begin{eqnarray}\label{equation Na2S condensation approximation}
10^{4}/T_{\textrm{\scriptsize{cond}}}(\textrm{Na}_{2}\textrm{S})& \approx &10.05-0.72(\log P_{T}+[\textrm{Na/H}]+0.5[\textrm{S/H}])\nonumber\\
& \approx &10.05-0.72\log P_{T}-1.08[\textrm{M/H}],
\end{eqnarray}
assuming [Na/H] = [S/H] (see $\S$\ref{s Computational Method}), and is shown in Figure \ref{figure sulfur chemistry} for a solar-metallicity gas.
Below the Na$_{2}$S cloud, monatomic Na is the most abundant Na-bearing gas, followed closely by NaCl.  With decreasing temperatures, NaCl becomes
increasingly important, but its abundance never exceeds that of monatomic Na below the cloud \citep{lodders 1999a}. Sodium sulfide condensation
effectively removes all sodium and $\sim6.5\%$ of all sulfur from the atmosphere because the protosolar Na abundance is 13\% of the protosolar S
abundance.

\subsubsection{Sphalerite, ZnS, condensation}\label{sss Zinc Sulfide, ZnS}

Sphalerite clouds form by the net thermochemical reaction
\begin{equation}\label{equation ZnS condensation reaction}
\textrm{Zn}+\textrm{H}_{2}\textrm{S}=\textrm{ZnS(s)}+\textrm{H}_{2}.
\end{equation}
The condensation temperature of ZnS as a function of $P_{T}$ and metallicity is given by
\begin{eqnarray}\label{equation ZnS condensation approximation}
10^{4}/T_{\textrm{\scriptsize{cond}}}(\textrm{ZnS})& \approx &12.52-0.63(\log P_{T}+[\textrm{Zn/H}]+[\textrm{S/H}])\nonumber\\
& \approx &12.52-0.63\log P_{T}-1.26[\textrm{M/H}],
\end{eqnarray}
assuming [Zn/H] = [S/H] (see $\S$\ref{s Computational Method}).  Sphalerite cloud formation effectively removes all Zn and $\sim0.3\%$ of H$_{2}$S
from the atmosphere.

\subsubsection{Ammonium Hydrogen Sulfide, NH$_{\textrm{\scriptsize{4}}}$SH, condensation}\label{sssAmmonium Hydrogen Sulfide, NH$_{4}$SH}
Sulfur condenses as NH$_{4}$SH via the net thermochemical reaction
\begin{equation}\label{equationNH4SHcondensation}
\textrm{NH}_{3}+\textrm{H}_{2}\textrm{S}=\textrm{NH}_{4}\textrm{SH(s)},
\end{equation}
in the cool upper atmospheres of giant planets (Figure \ref{figure sulfur chemistry}), where H$_{2}$S and NH$_{3}$ are the major S- and N-bearing
gases, respectively.  The NH$_{4}$SH condensation temperature as a function of $P_{T}$ and metallicity is approximated by
\begin{eqnarray}\label{equation NH4SH condensation approximation}
10^{4}/T_{\textrm{\scriptsize{cond}}}(\textrm{NH}_{4}\textrm{SH})& \approx &48.91-4.15(\log P_{T}+0.5[\textrm{N/H}]+0.5[\textrm{S/H}])\nonumber\\
& \approx &48.91-4.15\log P_{T}-4.15[\textrm{M/H}],
\end{eqnarray}
assuming [N/H] = [S/H] (see $\S$\ref{s Computational Method}).  This expression differs slightly from \citet{lodders and fegley 2002} due to the
revised solar elemental abundances of \citet{lodders 2003}.  The formation of a NH$_{4}$SH cloud is expected to efficiently remove all sulfur from
the atmosphere because NH$_{3}$ is $\sim4.4$ times more abundant than H$_{2}$S in a protosolar composition gas.

\subsubsection{Hydrogen Sulfide, H$_{\textrm{\scriptsize{2}}}$S, condensation}\label{sss Hydrogen Sulfide}
Hydrogen sulfide cloud formation is only possible in cool objects (such as Uranus and Neptune) if NH$_{3}$ is absent or less abundant than
H$_{2}$S, since ammonia will consume atmospheric sulfur to form NH$_{4}$SH via reaction (\ref{equationNH4SHcondensation}) \citep[e.g.,
see][]{fegley et al 1991}.  At sufficiently low temperatures, H$_{2}$S condensation may occur via the net reaction
\begin{equation}\label{equation H2S condensation reaction}
\textrm{H}_{2}\textrm{S}=\textrm{H}_{2}\textrm{S (s)}.
\end{equation}
The H$_{2}$S condensation temperature is approximated by
\begin{equation}\label{equation H2S condensation approximation}
10^{4}/T_{cond}(\textrm{H}_{2}\textrm{S}) \approx 86.49-8.54\log P_{T}-8.54[\textrm{S/H}].
\end{equation}
We found that no other S-bearing condensates are stable (OCS, SO$_{2}$, elemental S) over this \textit{P-T} range in a protosolar composition gas.

\subsection{Sulfur Chemistry in Substellar Objects}\label{ss Sulfur Chemistry in Substellar Objects}

\begin{figure}
\scalebox{.46}{\includegraphics{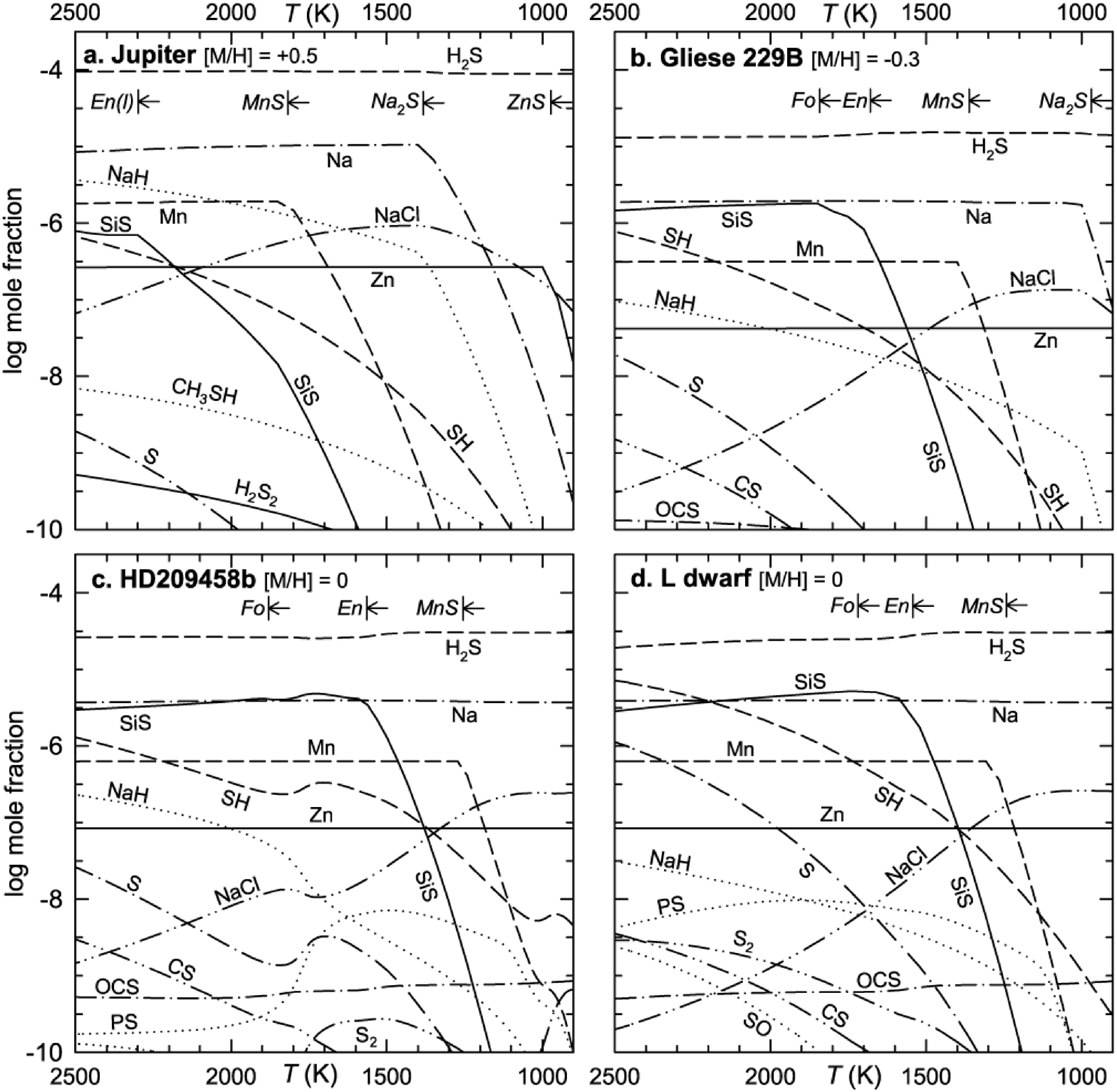}} \caption[Sulfur Chemistry Summary]{Sulfur chemistry along representative
atmospheric pressure-temperature profiles from 900 to 2500 K for (a) Jupiter ([M/H] $\approx$ +0.5), (b) Gliese
229B ([M/H] $\approx$ -0.3), (c) HD209458b ([M/H] $\approx$ 0), and (d) an L dwarf ([M/H] $\approx$ 0;
$T_{\textrm{\scriptsize{eff}}}$ = 1800 K), assuming uniform elemental enrichments and depletions.  Also shown are
the calculated condensation temperatures of forsterite (Fo; Mg$_{2}$SiO$_{4}$), enstatite (En; MgSiO$_{3}$), MnS,
Na$_{2}$S, and ZnS. With the exception of forsterite (not shown) and enstatite on Jupiter, all condensates shown
here condense as solids.} \label{figure sulfur chemistry summary}
\end{figure}

Figures \ref{figure sulfur chemistry summary}a-\ref{figure sulfur chemistry summary}d summarize the equilibrium sulfur gas chemistry computed
along the \textit{P-T} profiles of four representative substellar objects: Jupiter ($T_{\textrm{\scriptsize{eff}}}=124$ K), the T dwarf Gliese
229B ($T_{\textrm{\scriptsize{eff}}}=960$ K), the Pegasi planet HD209458b ($T_{\textrm{\scriptsize{eff}}}=1350$ K), and an L dwarf
($T_{\textrm{\scriptsize{eff}}}=1800$ K) (see $\S$\ref{ss Overview of Sulfur Chemistry}).  The profiles for Jupiter and Gliese 229B were computed
at $[\textrm{M/H}]\approx+0.5$ and $[\textrm{M/H}]\approx-0.3$, respectively, assuming uniform enrichments and depletions in elemental abundances.
The upper atmosphere of HD209458b is heated by a large incoming stellar flux, which is responsible for the shape of this planet's
\textit{\textit{P-T}} profile (Figure \ref{figure sulfur chemistry}) and hence the curvature in all gas abundances near 1800 K .  Also shown are
gases which are strongly affected by S-bearing cloud formation (e.g., Na, NaCl, Mn, Zn). The condensation temperatures of Mg$_{2}$SiO$_{4}$ (Fo),
MgSiO$_{3}$ (En), MnS, Na$_{2}$S, and ZnS are indicated by arrows (see Table \ref{table sulfur condensation}).  The Jupiter profile intersects the
forsterite condensation curve at high temperatures outside the range of Figure \ref{figure sulfur chemistry summary}a.  The Gliese 229B profile
intersects the ZnS condensation curve at 780 K, outside the range of Figure \ref{figure sulfur chemistry summary}b.  In contrast, ZnS does not
condense in the atmospheres of HD209458b or the L dwarf.

Hydrogen sulfide is the most abundant S-bearing gas throughout the atmospheres of substellar objects
($X_{\textrm{\scriptsize{H}}_{2}\textrm{\scriptsize{S}}}\approx X_{\Sigma\textrm{\scriptsize{S}}}$). The H$_{2}$S abundance slightly increases
when silicon in SiS is removed by Mg-silicate cloud formation (e.g., see Figures \ref{figure sulfur chemistry summary}b-\ref{figure sulfur
chemistry summary}d), and slightly decreases when sulfur is removed by metal sulfide cloud formation (e.g., see Figure \ref{figure sulfur
chemistry summary}a).

The \textit{Galileo} entry probe measured H$_{2}$S down to the $\sim$ 20 bar level on Jupiter \citep{niemann et al 1998, wong et al 2004}, where
the observed H$_{2}$S abundance of 2.4 times the solar H$_{2}$S/H$_{2}$ ratio should represent the total sulfur inventory in Jupiter's atmosphere
\citep[e.g.,][]{barshay and lewis 1978, fegley and lodders 1994, lodders 2004b}.  Ground-based observations have failed to detect H$_{2}$S on
Jupiter and Saturn because of its removal by condensation, its short photochemical lifetime, and the lack of major H$_{2}$S features in the 5
$\mu$m window. Observations of Jupiter at 2.7 $\mu$m by KAO \citep{larson et al 1984} and at 8.5 $\mu$m by \textit{Voyager} \citep{bezard et al
1983} provide only upper limits because these wavelengths probe atmospheric levels where H$_{2}$S is depleted by NH$_{4}$SH condensation.  The HST
detection of H$_{2}$S in the Shoemaker-Levy 9 impact region suggested excavation of Jovian atmosphere from regions below the NH$_{4}$SH cloud
where H$_{2}$S is more abundant \citep{noll et al 1995, zahnle et al 1995}.

Based on tropospheric NH$_{3}$ and H$_{2}$S abundances, NH$_{4}$SH cloud formation is expected to occur near the 220 K, 2.4 bar level on Jupiter
\citep[cf.][]{fegley and lodders 1994}.  This should efficiently remove H$_{2}$S from the gas because NH$_{3}$/H$_{2}$S $\sim7.5\pm3.4$ in
Jupiter's atmosphere \citep{wong et al 2004}.  The inferred NH$_{4}$SH cloud layer detected by the \textit{Galileo} entry probe nephelometer was
thin and tenuous, presumably due to the probe's entry in a relatively clear 5 $\mu$m hot spot \citep{west et al 2004}.

In contrast, NH$_{4}$SH cloud formation is not expected in the warmer atmospheres of brown dwarfs or Pegasi planets because their \textit{P-T}
profiles do not intersect the NH$_{4}$SH condensation curve (see Figure \ref{figure sulfur chemistry}), and H$_{2}$S gas remains. Measurements of
H$_{2}$S in these objects should therefore approximately represent their bulk atmospheric sulfur inventory \citep[e.g.][]{fegley and lodders
1996}.  \citet{saumon et al 2000} investigated the observability of H$_{2}$S and found that a H$_{2}$S feature at 2.1 $\mu$m is potentially
detectable in the infrared spectrum of Gliese 229B.  In the uppermost atmospheres of brown dwarfs and Pegasi planets, H$_{2}$S is plausibly
destroyed by photochemistry (see $\S$\ref{s Photochemistry}), as is expected for Jupiter \citep[e.g.,][]{prinn and owen 1976}.

Over the \textit{P-T} range considered here, the next most abundant S-bearing gases after H$_{2}$S are SH (at low temperatures) or SiS (at high
temperatures). The mercapto radical is generally the second or third most abundant sulfur gas throughout substellar atmospheres. \citet{yamamura
et al 2000} identified SH at $\sim3.67$ $\mu$m in the published infrared spectrum of R Andromedae, a S-type star. Their inferred SH/H ratio of
$\sim10^{-7}$ is consistent with thermochemical equilibrium \citep[][and Figure \ref{figure individual S gas}b]{tsuji 1973}. \citet{berdyugina and
livingston 2002} identified SH at $\sim3280$ {\AA} in the Sun's photosphere. They do not give column densities, but computed synthetic spectra at
5250 and 4750 K. We calculated SH mole fractions of $\sim9\times10^{-9}$ (5250 K) and $\sim5\times10^{-9}$ (4750 K) using the solar model
atmosphere of \citet{edvardsson et al 1993}. Using the effective temperatures for reference, we also compute SH column densities of
$\sim7\times10^{12}$ (960 K), $\sim2\times10^{17}$ (1350 K), and $\sim2\times10^{17}$ cm$^{-2}$ (1800 K) on Gliese 229B, HD209458b, and the L
dwarf, respectively, versus the SH column density of $\sim4\times10^{20}$ cm$^{-2}$ in R Andromedae \citep{yamamura et al 2000}.  The equilibrium
SH abundances expected for brown dwarfs and Pegasi planets may be too low for reliable quantitative analysis.

The relative importance of SiS increases with increasing temperature, and peak SiS abundances of $\sim$ 6 ppm at [M/H] = 0 and $\sim$ 60 ppm at
[M/H] = +0.5 are achieved at \textit{P-T} conditions close to the condensation temperatures of the Mg-silicate clouds. With their condensation,
silicon is removed from the atmosphere and the abundance of SiS and other Si-bearing gases drop.  This makes SiS a potential tracer of weather in
Pegasi planets and L dwarfs, analogous to FeH in T dwarfs \citep{burgasser et al 2002}, and gaps in the Mg-silicate cloud layers may expose
regions where SiS is relatively abundant. In cooler objects, the SiS fundamental absorption band near 13.3 $\mu$m may be difficult to distinguish
from ammonia features as NH$_{3}$ replaces N$_{2}$ to become the dominant N-bearing gas; the first SiS overtone at 6.6 $\mu$m has been observed in
the carbon giant WZ Cas \citep{aoki et al 1998}, though this feature will likely be overwhelmed by H$_{2}$O absorption in substellar
atmospheres.

The increasing importance of NaCl instead of Na and the removal of Na by Na$_{2}$S cloud formation is a plausible cause of the observed weakening
of Na atomic lines throughout the L dwarf spectral sequence and their disappearance in early T dwarfs \citep[e.g.][]{kirkpatrick et al 1999,
mclean et al 2003}; this is further enhanced by pressure broadening of the Na I doublet \citep[e.g.][]{tsuji et al 1996, burrows et al 2000b,
liebert et al 2000}. Sodium sulfide is expected to condense at the $\sim 1370$ K level in Jupiter's atmosphere and the $\sim 1000$ K level in
Gliese 229B (see Table \ref{table sulfur condensation}).  The L dwarf and the Pegasi planet HD209458b are too warm for Na$_{2}$S condensation
(assuming [M/H] $\approx$ 0; see Figure \ref{figure sulfur chemistry}) and Na remains in the gas.  This is consistent with the detection of Na in
the atmosphere of HD209458b \citep{charbonneau et al 2002}, though the chemistry results of \citet{lodders 1999a} suggest that limited Na$_{2}$S
cloud formation may occur on this planet if night-side temperatures are low enough \citep[e.g.,][]{fortney et al 2003, fortney et al 2005, iro et
al 2005, barman et al 2005}, or if its metallicity is sufficiently enhanced. However, while Na chemistry is strongly affected by Na$_{2}$S
condensation, S chemistry remains almost unchanged because Na$_{2}$S cloud formation only removes 6\% of all sulfur (see $\S$\ref{sss Sodium
Sulfide}).

The formation of the MnS cloud layer effectively removes Mn from the atmospheres of substellar objects. Similarly, ZnS condensation removes Zn
from the atmosphere of Jupiter ([M/H] $\approx$ +0.5) and Gliese 229B ([M/H] $\approx$ -0.3) above the $\sim$ 980 K and $\sim$ 780 K levels,
respectively.  In contrast, the atmospheres of L dwarfs and Pegasi planets are too warm for ZnS condensation and Zn will remain in the gas.

\section{Phosphorus Chemistry}\label{s Phosphorus Chemistry}

\subsection{Phosphorus Gas Chemistry}\label{ss Overview of Phosphorus Gas Chemistry}

The phosphorus equilibrium gas chemistry as a function of temperature and total pressure in a protosolar composition gas is illustrated in Figure
\ref{figure phosphorus chemistry}. The \textit{P-T} regions indicating the most abundant phosphorus bearing gas are bounded by solid lines. Also
shown is the condensation curve for NH$_{4}$H$_{2}$PO$_{4}$ (dotted line), model atmosphere profiles for representative substellar objects (dashed
lines; see $\S$\ref{ss Overview of Sulfur Chemistry}), and the H$_{2}$=H and CH$_{4}$=CO equal abundance curves (dash-dot lines). Phosphorus
chemistry is considerably more complex than that for C, N, O, and S, and several P-bearing species become abundant at elevated temperatures
\citep{barshay and lewis 1978, fegley and lewis 1980, fegley and lodders 1994}.

Under equilibrium conditions, P$_{4}$O$_{6}$ is the dominant phosphorus gas at low temperatures, and is replaced by PH$_{3}$ or P$_{2}$ as
temperatures increase.  The conversion of P$_{4}$O$_{6}$ to PH$_{3}$ occurs by net thermochemical reaction:
\begin{equation}\label{reaction PH3:P4O6}
\textrm{P}_{4}\textrm{O}_{6}+12\textrm{H}_{2}=4\textrm{PH}_{3}+6\textrm{H}_{2}\textrm{O}.
\end{equation}
In CH$_{4}$-dominated objects, the PH$_{3}$=P$_{4}$O$_{6}$ equal abundance boundary is approximated by
\begin{eqnarray}\label{equation line PH3:P4O6 CH4}
\log P_{T}&\approx& -11.40 + 12360/T + 2[\textrm{O/H}] + [\textrm{P/H}]\nonumber\\
&\approx& -11.40 + 12360/T + 3[\textrm{M/H}],
\end{eqnarray}
assuming [O/H] $\approx$ [P/H] (see $\S$\ref{s Computational Method}).  The net reaction of P$_{4}$O$_{6}$ to P$_{2}$ occurs at lower $P_{T}$ in
CO-dominated objects:
\begin{equation}\label{reaction P2:P4O6}
\textrm{P}_{4}\textrm{O}_{6}+6\textrm{H}_{2}=2\textrm{P}_{2}+6\textrm{H}_{2}\textrm{O},
\end{equation}
and the position of the P$_{2}$=P$_{4}$O$_{6}$ equal abundance boundary is approximated by
\begin{eqnarray}\label{equation line P2:P4O6}
\log P_{T}&\approx& 45.17 - 51048/T - 6[\textrm{O/H}] - [\textrm{P/H}]\nonumber\\
&\approx& 45.17 - 51048/T - 7[\textrm{M/H}].
\end{eqnarray}
assuming [O/H] $\approx$ [P/H] (see $\S$\ref{s Computational Method}).  The relative importance of reactions (\ref{reaction PH3:P4O6}) and
(\ref{reaction P2:P4O6}) depends on the position of an object's \textit{P-T} profile relative to the PH$_{3}$-P$_{4}$O$_{6}$-P$_{2}$ triple point
at $T\sim1101$ K and $P_{T}\sim10^{-1.20}$ bar, where all three gases have equal abundances [A(PH$_{3}$) = A(P$_{4}$O$_{6}$) = A(P$_{2}$)
$\approx$ $\frac{1}{7}\Sigma$P]. The positions of all the triple points shown in Figure \ref{figure phosphorus chemistry} are listed in Table
\ref{table phosphorus triple points}.

In L dwarf atmospheres, PH$_{3}$ and P$_{2}$ have similar abundances and are converted into each other by the net thermochemical reaction
\begin{equation}\label{reaction PH3:P2}
2\textrm{PH}_{3}=\textrm{P}_{2}+3\textrm{H}_{2}.
\end{equation}
The PH$_{3}$=P$_{2}$ equal abundance boundary in a solar metallicity gas is approximated by
\begin{equation}\label{equation line PH3:P2}
\log P_{T}\approx -1.56\times10^{7}/T^{2}+2.13\times10^{4}/T-7.68,
\end{equation}
from 1101 to 1330 K, and shifts to higher $P_{T}$ and lower $T$ with increasing metallicity.  At deeper atmospheric levels, PH$_{2}$ becomes the
most abundant phosphorus gas. In T dwarfs and cool L dwarfs PH$_{2}$ replaces PH$_{3}$ via
\begin{equation}\label{reaction PH3:PH2}
\textrm{PH}_{3}=\textrm{PH}_{2}+0.5\textrm{H}_{2}.
\end{equation}
The PH$_{3}$=PH$_{2}$ boundary is given by
\begin{equation}\label{equation line PH3:PH2}
\log P_{T} \approx 7.71 - 10888/T,
\end{equation}
and is independent of metallicity because the $m$ dependence for each P-bearing gas in reaction (\ref{reaction PH3:PH2}) cancels and the H$_{2}$
abundance is essentially constant over small metallicity variations.

In Pegasi planets and hot L dwarfs, PH$_{2}$ replaces P$_{2}$ via
\begin{equation}\label{reaction P2:PH2}
\textrm{P}_{2}+2\textrm{H}_{2}=2\textrm{PH}_{2}.
\end{equation}
The position of the P$_{2}$=PH$_{2}$ boundary in a solar metallicity gas is approximated by
\begin{equation}\label{equation line PH2:P2}
\log P_{T}\approx -1.68\times10^{7}/T^{2}+2.62\times10^{4}/T-10.68.
\end{equation}
from 1330 to 1780 K, and shifts to higher $P_{T}$ and $T$ with increasing metallicity.  The PH$_{3}$=P$_{2}$, PH$_{3}$=PH$_{2}$, and
P$_{2}$=PH$_{2}$ boundaries intersect at $T\sim 1330$ K and $P_{T}\sim10^{-0.48}$ bar, where all three gases have equal abundances [A(PH$_{3}$) =
A(PH$_{2}$) = A(P$_{2}$) $\approx$ $\frac{1}{4}\Sigma$P].

At even higher temperatures, PH$_{2}$ thermally dissociates via
\begin{equation}\label{reaction PH2:P}
\textrm{PH}_{2}=\textrm{P}+\textrm{H}_{2},
\end{equation}
and monatomic P becomes the major P-bearing gas.  The position of the PH$_{2}$=P boundary given by
\begin{equation}\label{equation line PH2:P}
\log P_{T} \approx 5.00 - 11233/T,
\end{equation}
is independent of metallicity because the $m$ dependence for each P-bearing gas  cancels out in reaction (\ref{reaction PH2:P}).  As for sulfur,
thermal ionization of P only becomes important at temperatures higher than those shown in Figure \ref{figure phosphorus chemistry}, e.g.,
$\textrm{P}^{+}/\textrm{P}\sim1$ at 6088 K ($10^{-2}$ bars), 4836 K ($10^{-4}$ bars), and 4054 K ($10^{-6}$ bars).

\subsubsection{Tetraphosphorus Hexaoxide, P$_{\textrm{\scriptsize{4}}}$O$_{\textrm{\scriptsize{6}}}$ gas}\label{sss Tetraphosphorus Hexaoxide, P4O6}

The equilibrium abundances of P$_{4}$O$_{6}$ as a function of $T$ and $P_{T}$ for solar metallicity are illustrated in Figure \ref{figure
individual P gas}a.   At high temperatures ($\gtrsim1000$ K), the P$_{4}$O$_{6}$ abundances depend on which P-bearing gas is dominant. For
example, within the PH$_{3}$ field, the P$_{4}$O$_{6}$ abundance is controlled by reaction (\ref{reaction PH3:P4O6}) and is proportional to
$P_{T}^{-3}$ and $m^{10}$, assuming [O/H] $\approx$ [P/H] (see $\S$\ref{s Computational Method}). Inside the P$_{2}$ field, the P$_{4}$O$_{6}$
abundance is controlled by reaction (\ref{reaction P2:P4O6}) and is proportional to $P_{T}$ and $m^{8}$, assuming [O/H] $\approx$ [P/H] (see
$\S$\ref{s Computational Method}).  In both cases, the P$_{4}$O$_{6}$ abundance rapidly decreases with temperature and is strongly dependent on
metallicity.

At low temperatures ($\lesssim1000$ K), P$_{4}$O$_{6}$ is the dominant P-bearing gas at equilibrium (Figure \ref{figure phosphorus chemistry}),
and its abundance is proportional to $m$, assuming [O/H] $\approx$ [P/H] (see $\S$\ref{s Computational Method}). The maximum P$_{4}$O$_{6}$
abundance is $X_{\textrm{\scriptsize{P}}_{4}\textrm{\scriptsize{O}}_{6}}\approx \frac{1}{4} X_{\Sigma\textrm{\scriptsize{P}}}$ because each
molecule of P$_{4}$O$_{6}$ contains four phosphorus atoms.  However, completely insignificant amounts of P$_{4}$O$_{6}$ are expected in the upper
atmospheres of giant planets and T dwarfs because the PH$_{3}$ to P$_{4}$O$_{6}$ conversion is kinetically inhibited inside the PH$_{3}$ stability
field where essentially no P$_{4}$O$_{6}$ exists (see $\S$\ref{ss Phosphorus Chemistry in Substellar Objects}).

\subsubsection{Phosphine, PH$_{\textrm{\scriptsize{3}}}$ gas}\label{sss Phosphine}

Chemical equilibrium abundances of PH$_{3}$ are shown in Figure \ref{figure individual P gas}b as a function of $T$ and $P_{T}$.  At high
temperatures ($T\gtrsim1000$ K), PH$_{3}$ abundances generally increase with $P_{T}$ but decrease with $T$ when other P-bearing gases are
dominant. For example, within the P$_{2}$ field the PH$_{3}$ abundance is governed by reaction (\ref{reaction PH3:P2}) and is approximated by
\begin{equation}\label{equation PH3 in P2}
\log X_{\textrm{\scriptsize{PH}}_{3}}\approx -9.03+3737/T+\log P_{T}+0.5[\textrm{P/H}],
\end{equation}
proportional to $P_{T}$ and $m^{0.5}$.  At higher total pressures, PH$_{3}$ is the dominant P-bearing gas (see Figure \ref{figure phosphorus
chemistry}), and its abundance is proportional to $m$.  The maximum PH$_{3}$ abundance is $X_{\textrm{\scriptsize{PH}}_{3}}\approx
X_{\Sigma\textrm{\scriptsize{P}}}$.  Phosphine therefore approximately represents the total phosphorus inventory in the deep atmospheres of giant
planets and T dwarfs.  Furthermore, rapid vertical mixing and phosphine quenching from deep atmospheric regions where PH$_{3}$ dominates is
expected to give disequilibrium abundances of $X_{\textrm{\scriptsize{PH}}_{3}}\approx X_{\Sigma\textrm{\scriptsize{P}}}$ in the observable
atmospheres of giant planets and T dwarfs (see $\S$\ref{ss Phosphorus Chemistry in Substellar Objects}).

\subsubsection{Diatomic Phosphorus, P$_{\textrm{\scriptsize{2}}}$ gas}\label{sss Diatomic
Phosphorus}

Mole fraction contours for P$_{2}$ are given in Figure \ref{figure individual P gas}c as a function of $T$ and $P_{T}$.  In cooler objects (e.g.,
giant planets and T dwarfs), P$_{2}$ is governed by reaction (\ref{reaction PH3:P2}), and the mole fraction abundance of P$_{2}$ within the
PH$_{3}$ field is
\begin{equation}\label{equation P2 in PH3}
\log X_{\textrm{\scriptsize{P}}_{2}} \approx -1.01 - 7474/T-2\log P_{T}+2[\textrm{P/H}],
\end{equation}
proportional to $P_{T}^{-2}$ and $m^{2}$.  At lower total pressures ($\lesssim 10^{-0.48}$ bar), P$_{2}$ is the dominant P-bearing gas (see Figure
\ref{figure phosphorus chemistry}), and its abundance is proportional to $m$.  The maximum P$_{2}$ abundance is
$X_{\textrm{\scriptsize{P}}_{2}}\approx \frac{1}{2}X_{\Sigma\textrm{\scriptsize{P}}}$ because each molecule of P$_{2}$ contains two phosphorus
atoms.  With increasing temperatures, P$_{2}$ is replaced by PH$_{2}$ via reaction (\ref{reaction P2:PH2}) in the atmospheres of L dwarfs and
Pegasi planets.

\subsubsection{Phosphino, PH$_{\textrm{\scriptsize{2}}}$ gas}\label{sss Phosphino, PH$_{2}$}

Figure \ref{figure individual P gas}d illustrates the abundances of the PH$_{2}$ radical, which is the most abundant P-bearing gas in the deep
atmospheres of brown dwarfs and Pegasi planets.  At lower temperatures, the PH$_{2}$ abundance within the PH$_{3}$ field is governed by reaction
(\ref{reaction PH3:PH2}) and is given by
\begin{equation}\label{equation PH2 in PH3}
\log X_{\textrm{\scriptsize{PH}}_{2}}\approx -2.40-5444/T-0.5\log P_{T}+[\textrm{P/H}],
\end{equation}
proportional to $P_{T}^{-0.5}$ and $m$. When P$_{2}$ is the dominant phosphorus gas, the PH$_{2}$ abundance is governed by reaction (\ref{reaction
P2:PH2}) and is given by
\begin{equation}\label{equation PH2 in P2}
\log X_{\textrm{\scriptsize{PH}}_{2}}\approx -5.19-1707/T+0.5\log P_{T}+0.5[\textrm{P/H}],
\end{equation}
proportional to $P_{T}^{0.5}$ and $m^{0.5}$.  Equations (\ref{equation PH2 in PH3}) and (\ref{equation PH2 in P2}) show that the PH$_{2}$
abundances decrease with $P_{T}$ when PH$_{3}$ dominates, and increase with $P_{T}$ when P$_{2}$ dominates.  This explains the curvature of the
PH$_{2}$ mole fraction contours in Figure \ref{figure individual P gas}d.

\subsection{Phosphorus Condensation Chemistry}\label{ss Phosphorus Condensation Chemistry}
At low temperatures (Figure \ref{figure phosphorus chemistry}), equilibrium condensation of NH$_{4}$H$_{2}$PO$_{4}$ is expected to occur via the
net thermochemical reaction
\begin{equation}\label{NH4H2PO$ condensation reaction}
10\textrm{H}_{2}\textrm{O} + 4\textrm{NH}_{3} + \textrm{P}_{4}\textrm{O}_{6}=4\textrm{NH}_{4}\textrm{H}_{2}\textrm{PO}_{4} + 4\textrm{H}_{2}
\end{equation}
The condensation temperature of NH$_{4}$H$_{2}$PO$_{4}$ as a function of $P_{T}$ and metallicity is approximated by
\begin{eqnarray}\label{equation NH4H2PO4 condensation approximation}
10^{4}/T_{\textrm{\scriptsize{cond}}}(\textrm{NH}_{4}\textrm{H}_{2}\textrm{PO}_{4}) &\approx& 29.99-0.20\left(11\log P_{T}+10[\textrm{O/H}]+4[\textrm{N/H}]+1[\textrm{P/H}]\right)\nonumber\\
10^{4}/T_{\textrm{\scriptsize{cond}}}(\textrm{NH}_{4}\textrm{H}_{2}\textrm{PO}_{4}) &\approx& 29.99-0.20\left(11\log
P_{T}+15[\textrm{M/H}]\right),
\end{eqnarray}
assuming [O/H] = [N/H] = [P/H] (see $\S$\ref{s Computational Method}). If equilibrium were attained (which is not the case for the giant planets
in our Solar System), NH$_{4}$H$_{2}$PO$_{4}$ condensation would efficiently remove phosphorus from the upper atmospheres of giant planets and the
coolest T dwarfs because oxygen and nitrogen abundances are typically much larger than that of phosphorus. We found that neither P$_{4}$O$_{10}$
nor elemental P condense under equilibrium conditions over this \textit{P-T} range from a solar system composition gas.

\subsection{Phosphorus Chemistry in Substellar Objects}\label{ss
Phosphorus Chemistry in Substellar Objects}

\begin{figure}
\scalebox{.46}{\includegraphics{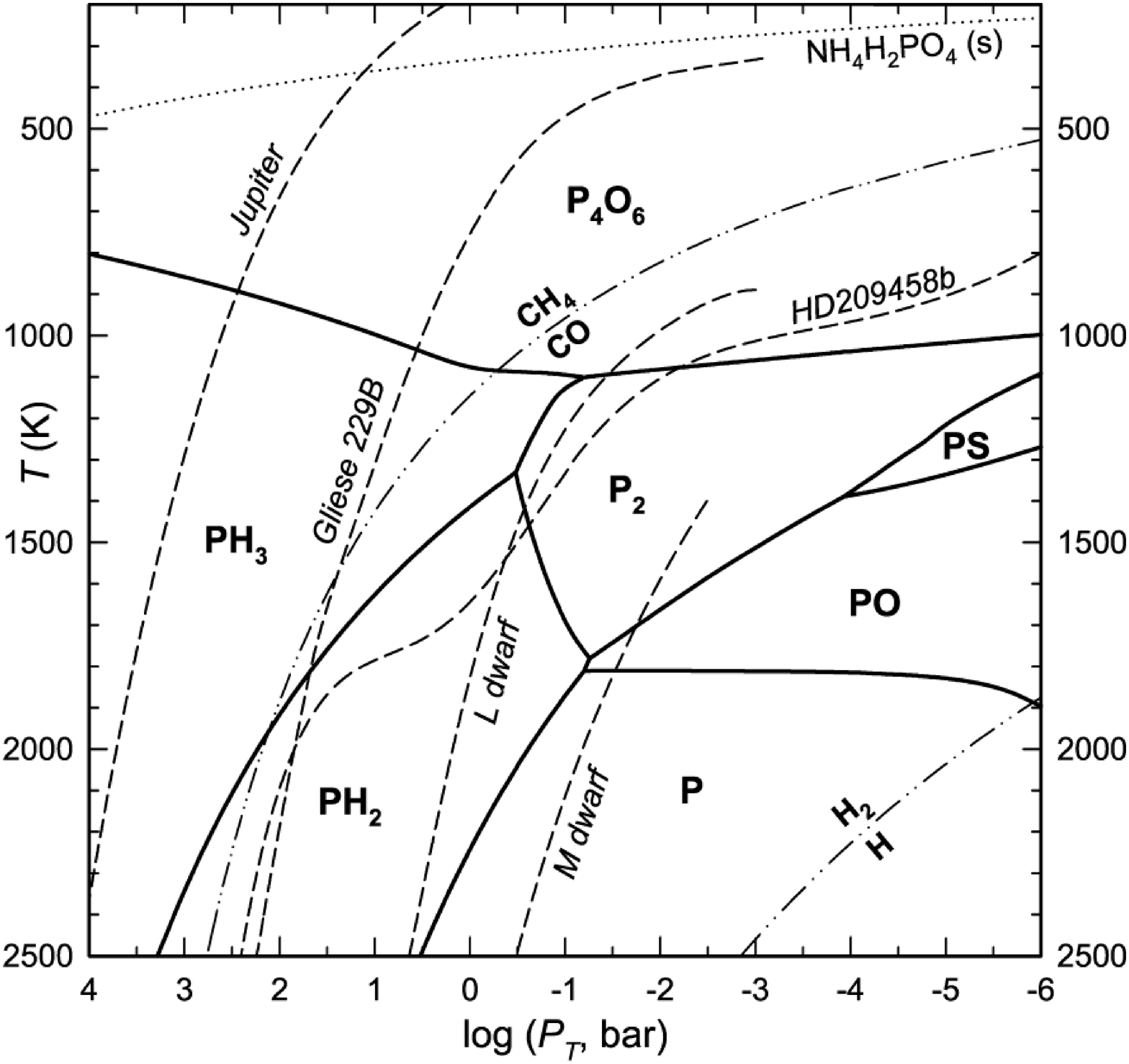}} \caption[Major P-bearing gases in a solar system composition
gas]{Overview of phosphorus chemistry as a function of temperature and pressure in a solar system composition gas.
The solid lines indicate where major P-bearing gases have equal abundances.  Also shown is the condensation curve
for NH$_{4}$H$_{2}$PO$_{4}$ (dotted line) and the position of the H$_{2}$=H equal abundance boundary (dash-dot
line). Atmospheric profiles for representative substellar objects are shown for reference (dashed lines).  See
text for details.} \label{figure phosphorus chemistry}
\end{figure}

\begin{figure}
\scalebox{.43}{\includegraphics{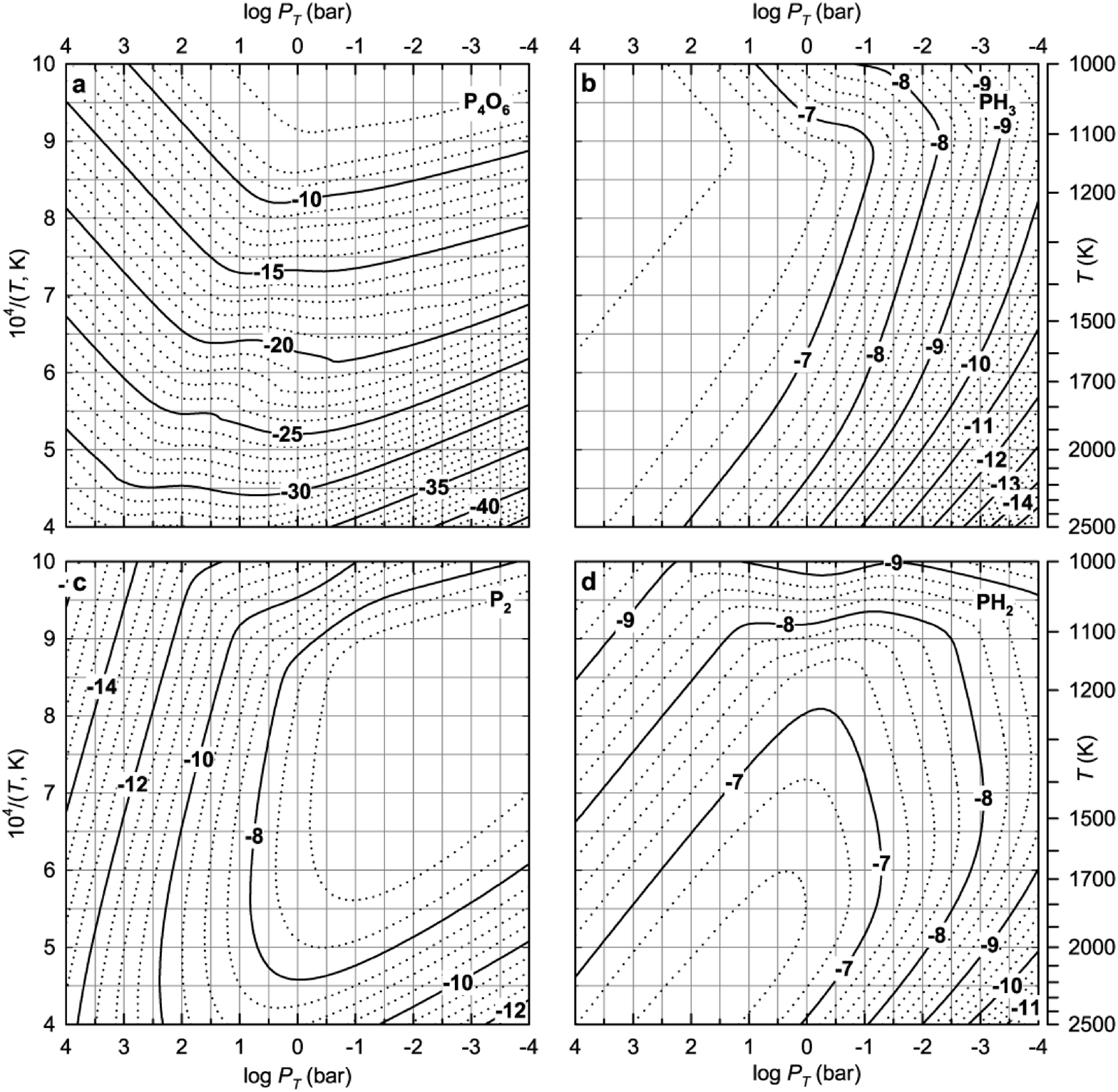}} \caption[Individual phosphorus gases]{Mole fraction contours (on a
logarithmic scale) for (a) tetraphosphorus hexaoxide (P$_{4}$O$_{6}$), (b) phosphine (PH$_{3}$), (c) diatomic
phosphorus (P$_{2}$), and (d) phosphino (PH$_{2}$) as a function of pressure and temperature in a solar
metallicity gas.  Gas abundances at higher or lower metallicities are given by the abundance expressions and/or
the dependence on the metallicity factor $m$.} \label{figure individual P gas}
\end{figure}

Figures \ref{figure phosphorus chemistry summary}a-\ref{figure phosphorus chemistry summary}d illustrate equilibrium phosphorus gas chemistry
along model atmosphere profiles Jupiter, Gliese 229B, HD209458b, and the L dwarf as done for sulfur in Figure \ref{figure sulfur chemistry
summary}.  As before, the shape of the \textit{P-T} profile of HD209458b (Figure \ref{figure phosphorus chemistry}) is responsible for the
curvature in the phosphorus abundances near 1800 K.

Using the effective temperatures for reference, the major P-bearing gases expected from thermochemical equilibrium in the observable atmospheres
are: P$_{4}$O$_{6}$ in giant planets and cool T dwarfs ($T_{\textrm{\scriptsize{eff}}}\sim100-1000$ K), PH$_{3}$ in hot T dwarfs and cool L dwarfs
($T_{\textrm{\scriptsize{eff}}}\sim1000-1400$ K), PH$_{2}$ in hot L dwarfs ($T_{\textrm{\scriptsize{eff}}}\sim1400-2000$ K), P$_{2}$ in Pegasi
planets such as HD209458b ($T_{\textrm{\scriptsize{eff}}}\sim1350$ K), and monatomic P in cool M dwarfs (e.g.,
$T_{\textrm{\scriptsize{eff}}}\sim2200$ K).  The dominant P-bearing gas changes with increasing temperature, and the major phosphorus gases at
depth are:  PH$_{2}$ in Pegasi planets, PH$_{2}$ and P in L dwarfs, PH$_{3}$ and PH$_{2}$ in T dwarfs, and PH$_{3}$ in giant planets (Figure
\ref{figure phosphorus chemistry}).

However, thermochemical equilibrium does not apply to phosphorus chemistry in the upper atmospheres of giant planets and T dwarfs because the
conversion from PH$_{3}$ to P$_{4}$O$_{6}$ is kinetically inhibited \citep{prinn and owen 1976, barshay and lewis 1978, prinn et al 1984, fegley
and lodders 1994}, and timescales for the PH$_{3}$ to P$_{4}$O$_{6}$ conversion ($t_{\textrm{\scriptsize{chem}}}$) are larger than typical
convective mixing timescales ($t_{\textrm{\scriptsize{mix}}}$) in these atmospheres \citep{fegley and prinn 1985, fegley and lodders 1994, fegley
and lodders 1996}. The chemical lifetime of PH$_{3}$ as a function of $T$ and $P_{T}$ for solar metallicity is given in Figure \ref{figure
PH3lifetime}, using PH$_{3}$ destruction kinetics described by \citet{visscher and fegley 2005}.  The quench temperature (where
$t_{\textrm{\scriptsize{chem}}}$ = $t_{\textrm{\scriptsize{mix}}}$) for PH$_{3}$ destruction is expected to be \emph{inside} the PH$_{3}$ field in
the deep atmospheres of giant planets and T dwarfs. As a result, virtually no P$_{4}$O$_{6}$ forms and PH$_{3}$ is mixed upward into their
observable atmospheres at abundances approximately representative of the total atmospheric phosphorus inventory
($X_{\textrm{\scriptsize{PH}}_{3}}\approx X_{\Sigma\textrm{\scriptsize{P}}}$).  We therefore expect PH$_{3}$ (instead of P$_{4}$O$_{6}$) to be the
major P-bearing gas in giant planets and T dwarfs.

Indeed, spectroscopic observations of PH$_{3}$ in the upper atmospheres of Jupiter and Saturn indicate tropospheric phosphine abundances of 1.1
ppm by volume on Jupiter \citep[e.g.,][and references therein]{lodders 2004b} and 4.5 ppm by volume on Saturn \citep[e.g.,][and references
therein]{visscher and fegley 2005}. These measured PH$_{3}$ abundances are $\sim$ 30 orders of magnitude higher than expected from thermodynamic
equilibrium and imply rapid vertical mixing from deeper atmospheric levels where phosphine is more abundant.  We expect PH$_{3}$ abundances of
$\sim 0.6$ ppm (for solar metallicity) in the upper atmospheres of Jupiter-type giant planets and cool T dwarfs such as Gliese 229B
\citep[e.g.,][]{fegley and lodders 1996}, and measurements of phosphine in these objects should approximately represent their bulk atmospheric
phosphorus inventory. \citet{noll and marley 1997} discussed the observability of gases affected by vertical mixing and concluded that the 4.3
$\mu$m feature of PH$_{3}$ is potentially detectable in the infrared spectra of T dwarfs. Photochemistry may plausibly destroy PH$_{3}$ in the
uppermost atmospheres of these objects (see $\S$\ref{s Photochemistry}), as occurs on Jupiter and Saturn.

In hotter objects, PH$_{3}$ may no longer represent the total atmospheric phosphorus inventory when P equilibrium chemistry is established and
other abundant P-bearing gases are present.  Phosphine abundances of $\sim50$ ppb (for solar metallicity) remain at the 1350 K level on HD209458b
and 1800 K in the model L dwarf. The PH$_{3}$ feature at 4.3 $\mu$m is potentially detectable in these objects if it can be distinguished from
nearby CO bands \citep{noll and marley 1997, saumon et al 2003}.

The abundant P-bearing gases at higher atmospheric temperatures include P$_{2}$, PH, PH$_{2}$, PH$_{3}$, and HCP (Figures \ref{figure phosphorus
chemistry summary}a-\ref{figure phosphorus chemistry summary}d).  The phosphino radical (PH$_{2}$) is the dominant P-bearing gas in the observable
atmosphere of the model L dwarf and in the deep atmospheres of T dwarfs and Pegasi planets.  However, PH$_{2}$ only represents about half of the
total atmospheric phosphorus inventory because other P-bearing gases remain relatively abundant within the PH$_{2}$ field. Equilibrium PH$_{2}$
abundances of $\sim1$ ppb, $\sim50$ ppb, and $\sim300$ ppb are expected (for solar metallicity) in the observable atmospheres of T dwarfs, Pegasi
planets, and L dwarfs, respectively.

As temperatures decrease, PH$_{2}$ is replaced by PH$_{3}$ (in Gliese 229B, Figure \ref{figure phosphorus chemistry summary}b) and/or P$_{2}$ (in
HD209458b and the L dwarf, Figures \ref{figure phosphorus chemistry summary}c-\ref{figure phosphorus chemistry summary}d) as the major P-bearing
gas, depending on the position of the atmospheric \textit{P-T} profile relative to the PH$_{3}$-PH$_{2}$-P$_{2}$ triple point (see Figure
\ref{figure phosphorus chemistry} and $\S$\ref{ss Overview of Phosphorus Gas Chemistry}).  Although P$_{2}$ is expected to be the dominant
P-bearing gas in the observable atmospheres of Pegasi planets ($X_{\textrm{\scriptsize{P}}_{2}}\sim0.2$ ppm at 1350 K) and the upper atmospheres
of L dwarfs, the P$_{2}$ bands expected near 0.2 $\mu$m \citep{carroll and mitchell 1975} may be very difficult to detect in these objects.  At
temperatures below $\sim1000$ K, P$_{2}$ is converted into P$_{4}$O$_{6}$, which is plausibly destroyed by photochemical processes in the
uppermost atmospheres of L dwarfs and Pegasi planets (see $\S$\ref{s Photochemistry}).

The abundances of PH and HCP are negligible in the observable atmospheres of giant planets and T dwarfs.  In warmer objects, these gases have
equilibrium abundances (at solar metallicity) of $\sim2$ ppb (PH) and $\sim20$ ppb (HCP) on HD209458b (at 1350 K), and $\sim50$ ppb (PH) and
$\sim15$ ppb (HCP) in the L dwarf model atmosphere (at 1800 K).

Under equilibrium conditions, NH$_{4}$H$_{2}$PO$_{4}$ cloud formation would be expected near the 400 K level in the upper atmosphere of Jupiter
\citep[cf.][]{barshay and lewis 1978, fegley and lodders 1994}, but is precluded by quenching of PH$_{3}$ (see above). Condensation of
NH$_{4}$H$_{2}$PO$_{4}$ is not expected in the warmer atmospheres of Gliese 229B, HD209458b, and the L dwarf because their \textit{P-T} profiles
do not intersect the NH$_{4}$H$_{2}$PO$_{4}$ condensation curve (Figure \ref{figure phosphorus chemistry}).

\section{Thermochemistry and Photochemistry}\label{s Photochemistry}

So far, our treatment of sulfur and phosphorus chemistry assumes the establishment of thermochemical equilibrium.  The exception is phosphine, for
which the kinetics of its destruction mechanism and convective vertical mixing must be considered for determining PH$_{3}$ abundances ($\S$\ref{ss
Phosphorus Chemistry in Substellar Objects}).  However, we must also consider photochemical reactions in the uppermost atmospheres of planets and
brown dwarfs that are companions to stars.  If ultraviolet flux causes photochemical reactions, the thermochemical reactions will be driven out of
equilibrium. Photochemistry may also play a role in isolated brown dwarfs if their coronae are warm enough to produce an ultraviolet flux
\citep{yelle 1999}.

For example, in Jupiter's troposphere the major C-, N-, S- and O-bearing gases expected from thermochemical equilibrium are CH$_{4}$, NH$_{3}$,
H$_{2}$S, and H$_{2}$O, respectively \citep{fegley and lodders 1994}.  However, Jupiter's stratosphere contains photochemically produced
hydrocarbons such as C$_{2}$H$_{6}$, C$_{2}$H$_{2}$, and C$_{2}$H$_{4}$ \citep{gladstone et al 1996}. Hence we must examine the relative roles of
thermochemistry and photochemistry on S and P compounds in brown dwarfs, EGPs, and related objects.

\subsection{Thermochemistry in Substellar Objects}\label{ss Thermochemistry in Substellar Objects}

Thermochemical equilibrium is readily achieved in the hot, deep atmospheres of substellar objects because gas phase reaction rates are generally
much faster than convective mixing rates (i.e., $t_{chem} \ll t_{mix}$).  To illustrate this, we consider the kinetics of the following elementary
reactions:
\begin{eqnarray}
\label{reaction H2S + H} \textrm{H} + \textrm{H}_{2}\textrm{S} &\rightarrow& \textrm{SH} + \textrm{H}_{2},\\
\label{reaction CO + SH} \textrm{SH} + \textrm{CO} &\rightarrow& \textrm{OCS} + \textrm{H},\\
\label{reaction H2S + S} \textrm{S} + \textrm{H}_{2}\textrm{S} &\rightarrow& \textrm{SH} + \textrm{SH},\\
\label{reaction Na + HCl} \textrm{Na} + \textrm{HCl} &\rightarrow& \textrm{NaCl} + \textrm{H},\\
\label{reaction PH2 + PH} \textrm{PH} + \textrm{PH}_{2} &\rightarrow& \textrm{PH}_{3} + \textrm{P},\\
\label{reaction PH3 + H} \textrm{H} + \textrm{PH}_{3} &\rightarrow& \textrm{PH}_{2} + \textrm{H}_{2}.
\end{eqnarray}
Here we use reaction (\ref{reaction H2S + H}) as an example for computing $t_{chem}$.  The rate of disappearance of H$_{2}$S in reaction
(\ref{reaction H2S + H}) is given by
\begin{equation}\label{equation H2S destruction}
\frac{d[\textrm{H}_{2}\textrm{S}]}{dt}=-k_{\ref{reaction H2S + H}}[\textrm{H}][\textrm{H}_{2}\textrm{S}],
\end{equation}
where $k_{\ref{reaction H2S + H}}$ (cm$^{3}$ molecules$^{-1}$ s$^{-1}$) is the rate coefficient for reaction (\ref{reaction H2S + H}) and
[H$_{2}$S] and [H] are the molecular number densities (molecules cm$^{-3}$) for H$_{2}$S and H.  The chemical lifetime for H$_{2}$S is defined as
\begin{equation}\label{equation H2S lifetime}
t_{chem}(\textrm{H}_{2}\textrm{S})=-\frac{[\textrm{H}_{2}\textrm{S}]}{d[\textrm{H}_{2}\textrm{S}]/dt}=\frac{1}{k_{\ref{reaction H2S +
H}}[\textrm{H}]},
\end{equation}
and is the time required for a 1/$e$ reduction in the H$_{2}$S abundance.  Representative chemical lifetimes for reactions (\ref{reaction H2S +
H})-(\ref{reaction PH3 + H}) at 1 bar total pressure and 1000, 1500, and 2000 K are listed in Table 1.  Although the short chemical lifetimes in
Table 1 indicate that these reactions proceed rapidly, we must also consider vertical mixing rates to determine if equilibrium is established. The
timescale for convective mixing, $t_{mix}$ is approximated by
\begin{equation}
t_{mix}\sim H^{2}/K_{eddy}
\end{equation}
where $K_{eddy}$ is the eddy diffusion coefficient.  The pressure scale height, $H$, is given by
\begin{equation}
H=RT/\mu g
\end{equation}
where $R$ is the gas constant, $\mu$ is the mean molecular weight of the gas, and $g$ is gravity.   For a solar system composition gas,
$\mu\approx 2.3$ g mol$^{-1}$.  A convective timescale of $t_{mix} \sim 10^{6}$ seconds is expected near the 1000 K level on Jupiter, using $H\sim
140$ km ($\log g = 3.4$) and $K_{eddy}\sim10^{8}$ cm$^{2}$ s$^{-1}$ \citep{fegley and prinn 1988}. On Gliese 229B, we estimate $t_{mix}\sim
10^{7}$ seconds at the 1000 K level, assuming $H\sim 3.6$ km \citep[$\log g=5.0$;][]{saumon et al 2000} and $K_{eddy}\sim 10^{4}$ cm$^{2}$
s$^{-1}$ \citep{griffith and yelle 1999}. Because of its low gravity \citep[$\log g=3.0$;][]{iro et al 2005} HD209458b has large mixing times of
$\sim 10^{7}$ to $10^{11}$ seconds at the 1000 K level, depending on the assumed $K_{eddy}$ value ($\sim 10^{8}$ to $10^{4}$ cm$^{2}$ s$^{-1}$).
These timescales for convective mixing are typically many orders of magnitude greater than the chemical lifetimes listed in Table \ref{Table
Representative Chemical Lifetimes} (i.e., $t_{chem} \ll t_{mix}$). We therefore expect the rapid establishment of thermochemical equilibrium in
the deep atmospheres of substellar objects. In instances where the rate of vertical mixing equals or exceeds the rate of chemical reactions
($t_{chem} \geq t_{mix}$), disequilibrium species (e.g., PH$_{3}$, see $\S$\ref{ss Phosphorus Chemistry in Substellar Objects}) are mixed into the
upper atmospheres of substellar objects, as occurs on Jupiter and Saturn \citep[e.g., CO, HCN, PH$_{3}$, GeH$_{4}$, AsH$_{3}$;][]{fegley and
lodders 1994} and Gliese 229B \citep[e.g. CO, PH$_{3}$, N$_{2}$;][]{fegley and lodders 1996, griffith and yelle 1999, saumon et al 2000}.

\subsection{Photochemistry on Pegasi Planets}\label{ss Photochemistry on Pegasi Planets}

In order to examine the relative importance of photochemical vs. thermochemical processes in substellar objects, we consider an extreme example:
the strongly irradiated upper atmosphere of HD209458b.  This Pegasi planet orbits a solar-type star at 0.05 AU and receives a stellar UV flux
$\sim$ 10,000 times that for Jupiter \citep[e.g.,][]{liang et al 2003}.  To determine the depth in the atmosphere where thermochemical processes
become important, we compared the thermochemical equilibrium abundances of the reactive species OH, O, and H in our model with photochemical
equilibrium abundances from \citet{liang et al 2003} for HD209458b. Plots comparing photochemical (dotted lines) and thermochemical (solid lines)
abundances for these reactive species are shown in Figure \ref{figure thermo vs photo}, using the \textit{P-T} profile for HD209458b from Figure 1
of \citet{liang et al 2003}.

At very low pressures, in the uppermost atmosphere of HD209458b, photochemistry produces excess disequilibrium abundances of OH, O, and H,
primarily via H$_{2}$O photolysis \citep{liang et al 2003}.  Thermochemistry becomes increasingly important with depth, as pressures and
temperatures increase and the available stellar flux decreases. Eventually, the predicted thermochemical abundances surpass the photochemical
abundances, indicating the establishment of equilibrium chemistry.  In other words, the photochemical reactions are no longer driving the
thermochemical reactions out of equilibrium.  As shown in Figure \ref{figure thermo vs photo}, this crossover occurs near the $\sim1$ to 10 mbar
levels ($T\sim980$ to 1140 K) for the reactive species OH, O, and H. Thermochemical processes are therefore expected to be important at pressures
of $\sim10$ mbar and higher on HD209458b, consistent with results showing that most of the incoming stellar flux is absorbed by the 1 bar level
($T\sim1640$ K) in its atmosphere \citep{iro et al 2005}. Furthermore, \citet{fortney et al 2005, fortney et al 2006} point out that the infrared
spectra of Pegasi planets appear to be most sensitive to opacity near 10 to 100 mbar ($T\sim1140$ to 1400 K). This suggests that thermochemical
models will provide a useful basis for interpreting and guiding observations of Pegasi planet atmospheres.

\section{Summary}\label{s Summary}

Thermochemistry governs the chemical behavior of sulfur and phosphorus species in the hot, deep atmospheres of substellar objects. Hydrogen
sulfide is the most abundant sulfur-bearing gas throughout substellar atmospheres, and observations of H$_{2}$S in these objects should provide a
good estimate of their atmospheric sulfur content. The condensation of metal sulfide clouds slightly lowers the atmospheric H$_{2}$S abundance to
$\sim91 \%$ of the total S inventory. These clouds also affect opacity by introducing condensed particles and by removing absorbing gases from the
observable atmosphere. Mercapto (SH) and SiS are the next most abundant sulfur gases after H$_{2}$S, and the relative importance of SiS increases
with increasing effective temperature. Furthermore, the maximum SiS abundance occurs near the Mg-silicate cloud base and is a potential tracer of
weather in Pegasi planets and L dwarfs.

Phosphorus speciation is considerably more complex than that for C, N, O, or S, and several phosphorus gases become relatively abundant at high
temperatures.  Disequilibrium abundances of PH$_{3}$ are expected in the upper atmospheres of giant planets and T dwarfs via rapid vertical mixing
from deeper levels.  In addition, phosphine abundances in giant planets and T dwarfs are expected to approximately represent their atmospheric
phosphorus inventories.  In Pegasi planets and L dwarfs, P$_{2}$ is the dominant phosphorus gas until replaced by P$_{4}$O$_{6}$ at low
temperatures or PH$_{2}$ at high temperatures. Phosphino (PH$_{2}$) is the most abundant phosphorus gas in the deep atmospheres of brown dwarfs
and Pegasi planets.

\acknowledgments This work was supported by the NASA Planetary Atmospheres Program.  Work by KL was supported by
NSF grant AST 04-06963.

\appendix
\section{Additional Compounds}
Here we list additional sulfur and phosphorus compounds investigated but not discussed in detail above.  This list includes unstable condensates
and minor gases predicted to have very low abundances in a solar system composition gas at temperatures and total pressures relevant for
substellar atmospheres. Depending on the prevailing sulfur, carbon, and phosphorus chemistry, the abundances of key gases are $P_{T}$ independent:
H$_{2}$S$_{2}$ within the H$_{2}$S field; OCS, CS, and CS$_{2}$ within the CO field; CH$_{3}$SH within the CH$_{4}$ field; HPO, PH, and PN within
the P$_{2}$ field. However, their abundances are generally too low to serve as useful temperature probes in brown dwarf or EGP atmospheres.  Some
gases (H$_{2}$S$_{2}$, CH$_{3}$SH) are potentially detectable via deep atmospheric entry probes on Jupiter and Saturn \citep{fegley and lodders
1994}.

For a given metallicity, relatively constant OCS abundances are expected throughout the atmospheres of CO-dominated objects; however, maximum OCS
abundances are only $\sim1$ ppb (for solar metallicity) in L dwarfs and Pegasi planets. The equilibrium abundances of P, PO, and PS generally
increase with $T$ and decrease with $P_{T}$.  These gases only become important at lowest total pressures considered here (see Figure \ref{figure
phosphorus chemistry}).  The behavior of HCP is similar to that of PH$_{2}$; maximum HCP abundances are $\sim50$ ppb (for solar metallicity) over
the \textit{P-T} range considered here.
\begin{center}\emph{Sulfur}\end{center}
S$_{3}$, S$_{4}$, S$_{2}$O, SO, SO$_{2}$, SO$_{3}$, SOH, HSO, H$_{2}$SO, H$_{2}$SO$_{4}$, SF, SF$_{2}$, H$_{2}$S$_{2}$, OCS, CS, CS$_{2}$,
CH$_{3}$SH, NS, PS, K$_{2}$S, CaS, FeS, MgS, MnS, TiS, VS, ZrS, SiS$_{2}$, TiS$_{2}$, ZrS$_{2}$, S(s,l), OCS(s,l), SO$_{2}$(s,l)
\begin{center}\emph{Phosphorus}\end{center}
P$_{3}$, P$_{4}$, PO, PO$_{2}$, P$_{2}$O$_{3}$, P$_{2}$O$_{4}$, P$_{2}$O$_{5}$, P$_{3}$O$_{6}$, P$_{4}$O$_{7}$, P$_{4}$O$_{8}$, P$_{4}$O$_{9}$,
P$_{4}$O$_{10}$, PCl, PCl$_{2}$, PCl$_{3}$, PF, PF$_{2}$, PF$_{3}$, HPO, HCP, CP, PH, PN, PS, P(s), P$_{4}$O$_{10}$(s,l)

\begin{deluxetable}{ccccc}
\tablewidth{0pt}\tablecolumns{5}\tablecaption{Equilibrium Condensation Temperatures of Sulfur-Bearing Compounds} \tablehead{\colhead{object} &
\colhead{MnS} & \colhead{Na$_{2}$S} & \colhead{ZnS} & \colhead{NH$_{4}$SH}}\startdata
Jupiter & 1800 & 1370 & 970 & 220\\
Gliese 229B & 1380 & 1000 & 780 & \nodata\\
HD209458b & 1260 & $\sim800$\tablenotemark{a} & \nodata & \nodata\\
L dwarf & 1270 & \nodata & \nodata & \nodata\\
M dwarf & \nodata & \nodata & \nodata & \nodata\\
\enddata
\tablecomments{Condensation temperatures computed using [M/H] $\approx$ +0.5 for Jupiter, [M/H] $\approx$ -0.3 for Gliese 229B, and [M/H]
$\approx$ 0 for HD209458b and the model L dwarf.} \tablenotetext{a}{night-side condensation} \label{table sulfur condensation}
\end{deluxetable}

\begin{deluxetable}{ccccc}
\tablewidth{0pt}\tablecolumns{5}\tablecaption{Phosphorus Equal Abundance ``Triple Points" in a Solar Metallicity Gas} \tablehead{\colhead{triple
point\tablenotemark{a}} & \colhead{$T$, K} & \colhead{$\log (P_{T},\textrm{bar})$}} \startdata
PH$_{3}$-P$_{4}$O$_{6}$-P$_{2}$ & 1101 & -1.20\\
PH$_{3}$-PH$_{2}$-P$_{2}$ & 1330 & -0.48\\
P$_{2}$-PO-PS & 1389 & -3.93\\
P$_{2}$-PH$_{2}$-PO & 1780 & -1.26\\
PH$_{2}$-P-PO & 1811 & -1.20\\
\enddata
\tablenotetext{a}{e.g., where $X_{\textrm{\scriptsize{A}}}=X_{\textrm{\scriptsize{B}}}=X_{\textrm{\scriptsize{C}}}$ for phosphorus-bearing gases
A, B, C.} \label{table phosphorus triple points}
\end{deluxetable}

\begin{deluxetable}{cccccl}
\tablewidth{0pt}\tablecolumns{6}\tablecaption{Representative Chemical Lifetimes at 1 bar pressure} \tablehead{ & & \multicolumn{3}{c}{chemical
lifetime, s} & \\ \colhead{reaction} & \colhead{$t_{chem}$} & \colhead{1000 K} & \colhead{1500 K} & \colhead{2000 K} & \colhead{kinetic
data}}\startdata
\ref{reaction H2S + H} & $t_{chem}(\textrm{H}_{2}\textrm{S})$ & $10^{0.4}$ & $10^{-3.8}$ & $10^{-5.9}$ & \citet{yoshimura et al 1992}\\
\ref{reaction CO + SH} & $t_{chem}(\textrm{SH})$ & $10^{2.7}$ & $10^{0.2}$ & $10^{-0.2}$ & \citet{kurbanov and mamedov 1995}\\
\ref{reaction H2S + S} & $t_{chem}(\textrm{S})$ & $10^{1.7}$ & $10^{-2.3}$ & $10^{-4.3}$ & \citet{woiki and roth 1994}\\
\ref{reaction Na + HCl} & $t_{chem}(\textrm{Na})$ & $10^{1.4}$ & $10^{-1.6}$ & $10^{-1.8}$ & \citet{husain and marshall 1986}\\
\ref{reaction PH2 + PH} & $t_{chem}(\textrm{PH})$ & $10^{4.2}$ & $10^{1.2}$ & $10^{1.0}$ & \citet{twarowski 1995}\\
\ref{reaction PH3 + H} & $t_{chem}(\textrm{PH}_{3})$ & $10^{0.4}$ & $10^{-3.5}$ & $10^{-5.4}$ & \citet{arthur and cooper 1997}\\
\enddata \label{Table Representative Chemical Lifetimes}
\tablecomments{Convective mixing timescales (in seconds) are $t_{\textrm{\scriptsize{conv}}}\sim$ $10^{6.3}$ (1000
K), $10^{6.6}$ (1500 K), and $10^{6.9}$ (2000 K) for Jupiter, $10^{7.1}$ (1000 K), $10^{7.5}$ (1500 K), and
$10^{7.7}$ (2000 K) for Gliese 229B, and $10^{7.1}$ to $10^{11.1}$ (1000 K), $10^{7.5}$ to $10^{11.5}$ (1500 K),
and $10^{7.7}$ to $10^{11.7}$ (2000 K) for HD209458b. See \S\ref{ss Thermochemistry in Substellar Objects} for
details.}
\end{deluxetable}

\clearpage

\begin{figure}
\scalebox{.46}{\includegraphics{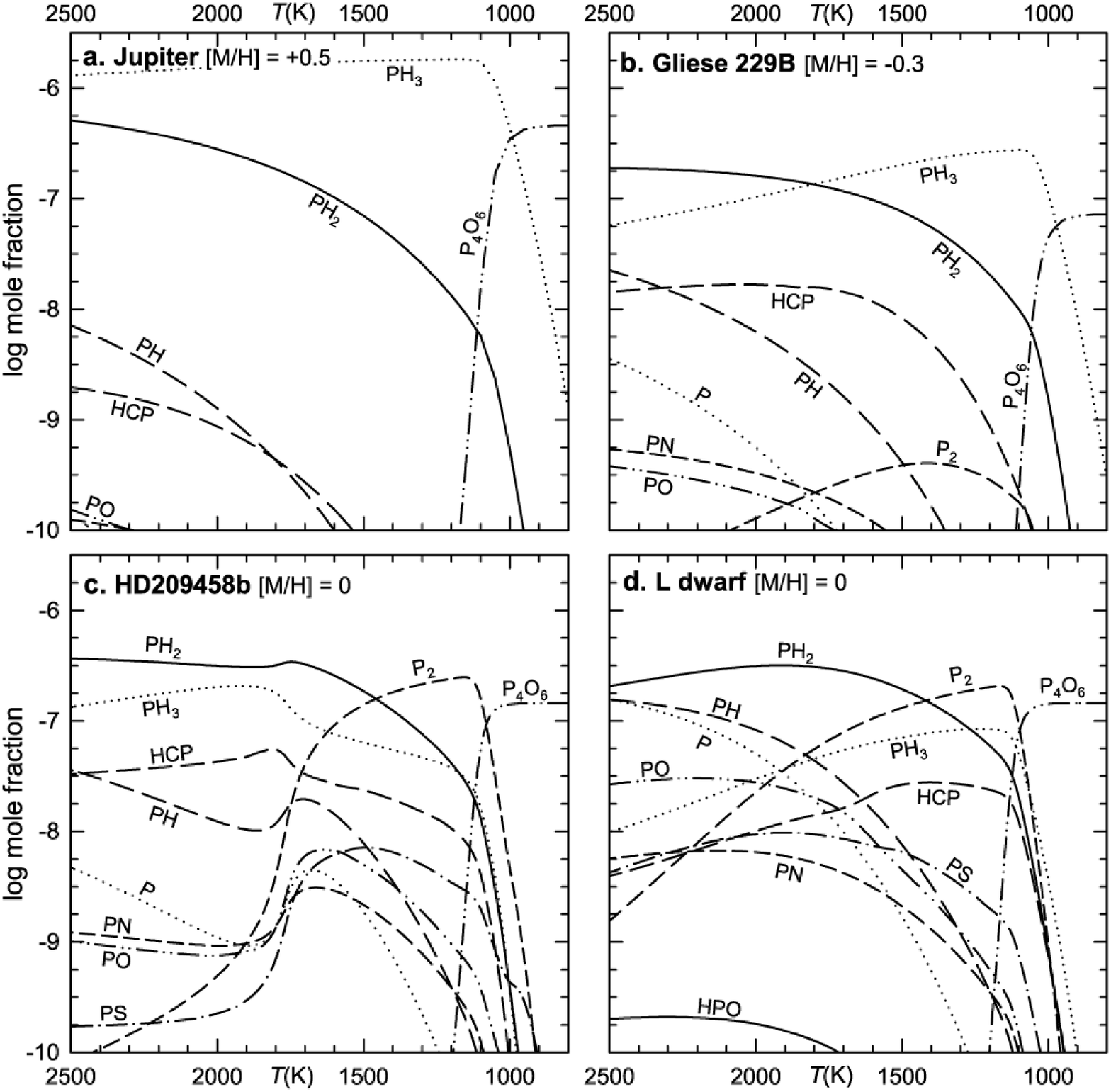}} \caption[Phosphorus Chemistry Summary]{Phosphorus equilibrium chemistry
along model atmosphere \textit{P-T} profiles from 800 to 2500 K for (a) Jupiter ([M/H] $\approx$ +0.5;
$T_{\textrm{\scriptsize{eff}}}$ = 124 K), (b) Gliese 229B ([M/H] $\approx$ -0.3; $T_{\textrm{\scriptsize{eff}}}$ =
960 K), (c) HD209458b ([M/H] $\approx$ 0; $T_{\textrm{\scriptsize{eff}}}$ = 1350 K), and (d) an L dwarf ([M/H]
$\approx$ 0; $T_{\textrm{\scriptsize{eff}}}$ = 1800 K), assuming uniform elemental enrichments and depletions.
Note that the PH$_{3}$ to P$_{4}$O$_{6}$ conversion is kinetically inhibited in giant planets and T dwarfs, and
that Jupiter's observable atmosphere lies outside the temperature range shown here.} \label{figure phosphorus
chemistry summary}
\end{figure}

\begin{figure}
\scalebox{.43}{\includegraphics{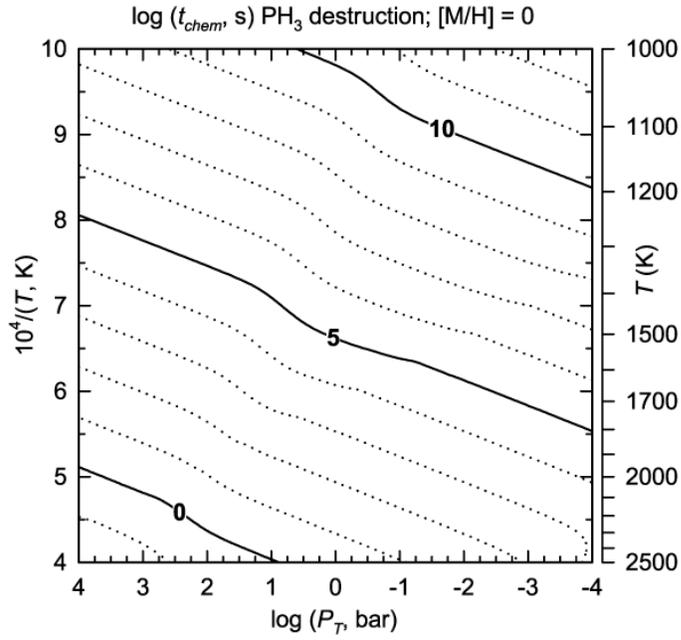}} \caption[Phosphine, PH$_{3}$ chemical lifetime]{Logarithmic time scale
(seconds) for the chemical conversion of PH$_{3}$ to P$_{4}$O$_{6}$ in a solar metallicity gas. This conversion is
kinetically inhibited in the upper atmospheres of giant planets and T dwarfs.} \label{figure PH3lifetime}
\end{figure}

\begin{figure}
\scalebox{.4}{\includegraphics{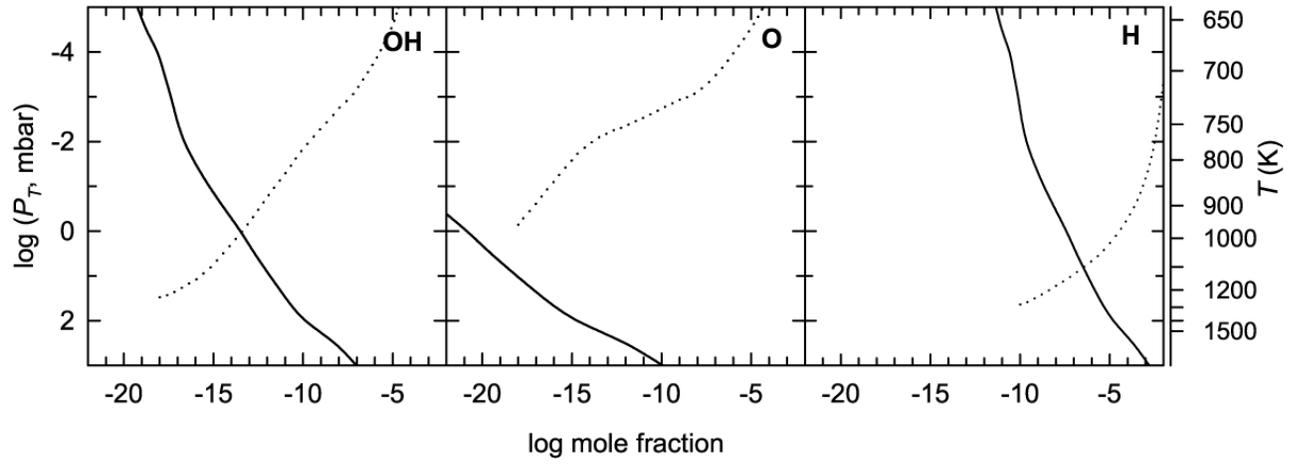}} \caption[Thermochemical vs Photochemical Abundances]{Comparison of
thermochemical (solid lines) and photochemical \citep[dotted lines;][]{liang et al 2003} abundances for the
reactive species OH, O, and H in the upper atmosphere of HD209458b. See text for details.} \label{figure thermo vs
photo}
\end{figure}


\begin{thebibliography}{}
\bibitem[Aoki et al.(1998)]{aoki et al 1998} Aoki, W., Tsuji, T., \& Ohnaka, K.\ 1998, \aap, 340, 222
\bibitem[Arthur \& Cooper(1997)]{arthur and cooper 1997} Arthur, N. L., \& Cooper, I. A. 1997, J. Chem. Soc. Faraday Trans., 93, 521
\bibitem[Barman et al.(2005)]{barman et al 2005} Barman, T. S., Hauschildt, P. H., \& Allard, F.\ 2005, \apj, 632, 1132
\bibitem[Barshay \& Lewis(1978)]{barshay and lewis 1978} Barshay, S. S., \& Lewis, J. S. 1978, \icarus, 33, 593
\bibitem[Baulch et al.(1992)]{baulch et al 1992} Baulch, D. L., et al. 1992, {J. Phys. Chem Ref. Data}, 21, 411
\bibitem[Berdyugina \& Livingston(2002)]{berdyugina and livingston 2002} Berdyugina, S.~V., \& Livingston, W.~C.\ 2002, \aap, 387, L6
\bibitem[B\'{e}zard et al.(1983)]{bezard et al 1983} B\'{e}zard, B., Marten, A., Baluteau, J. P., Gautier, D., Flaud, J. M., \& Camy-Peyret, C.\ 1983, Icarus, 55, 259
\bibitem[Burgasser et al.(2002)]{burgasser et al 2002} Burgasser, A. J., Marley, M. S., Ackerman, A. S., Saumon, D., Lodders, K., Dahn, C. C., Harris, H. C., \& Kirkpatrick, J. D. 2002, \apj, 571, L151
\bibitem[Burrows et al.(2000a)]{burrows et al 2000a} Burrows, A., Hubbard, W. B., Lunine, J. I., Marley, M. S. \& Saumon, D. 2000, in Protostars and Planets IV, ed. V. Mannings, A. P. Boss, \& S. S. Russell (Tuscon: Univ. of Arizona Press), 1339
\bibitem[Burrows et al.(2000b)]{burrows et al 2000b} Burrows, A., Marley, M. S., \& Sharp, C. M. 2000, \apj, 531, 438
\bibitem[Burrows et al.(2006)]{burrows et al 2006} Burrows, A., Sudarsky, D., \& Hubeny I. 2006, \apj, 640, 1063
\bibitem[Carroll \& Mitchell(1975)]{carroll and mitchell 1975} Carroll, P. K., \& Mitchell, P. I. 1975, Proc. Roy. Soc. Lon. A, 342, 93
\bibitem[Charbonneau et al.(2002)]{charbonneau et al 2002} Charbonneau, D., Brown, T. M., Noyes, R. W., \& Gilliland, R. L.\ 2002, \apj, 568, 377
\bibitem[Chase(1999)]{chase 1999} Chase, M. W. 1999, {J. Phys. Chem. Ref. Data}, 28, monograph no. 9
\bibitem[Chen et al.(2000)]{chen et al 2000} Chen, Y.~Q., Nissen, P.~E., Zhao, G., Zhang, H.~W., \& Benoni, T.\ 2000, \aaps, 141, 491
\bibitem[Edvardsson et al.(1993)]{edvardsson et al 1993} Edvardsson, B., Andersen, J., Gustafsson, B., Lambert, D. L., Nissen, P. E., \& Tomkin, J.\ 1993, \aap, 275, 101
\bibitem[Ecuvillon et al.(2004)]{ecuvillon et al 2004} Ecuvillon, A., Israelian, G., Santos, N. C., Mayor, M., Villar, V., \& Bihain, G.\ 2004, \aap, 426, 619
\bibitem[Fegley et al.(1991)]{fegley et al 1991} Fegley, B. Jr., Gautier, D., Owen, T., \& Prinn, R. G.\ 1991, in Uranus, ed. J. T. Bergstralh, E. D. Miner, \& M. S. Matthews (Tuscon: Univ. of Arizona Press), 147
\bibitem[Fegley \& Lewis(1979)]{fegley and lewis 1979} Fegley, B., Jr., \& Lewis, J. S. 1979, \icarus, 38, 166
\bibitem[Fegley \& Lewis(1980)]{fegley and lewis 1980} Fegley, B., Jr., \& Lewis, J. S. 1980, \icarus, 41, 439
\bibitem[FL94()Fegley \& Lodders 1994]{fegley and lodders 1994} Fegley, B., Jr., \& Lodders, K. 1994, \icarus, 110, 117; FL94
\bibitem[Fegley \& Lodders(1996)]{fegley and lodders 1996} Fegley, B., Jr., \& Lodders, K. 1996, \apj, 472, L37
\bibitem[Fegley \& Prinn(1985)]{fegley and prinn 1985} Fegley, B., Jr., \& Prinn, R. G. 1985, \apj, 299, 1067
\bibitem[Fegley \& Prinn(1988)]{fegley and prinn 1988} Fegley, B., Jr., \& Prinn, R. G. 1988, \apj, 324, 621
\bibitem[Fortney et al.(2003)]{fortney et al 2003} Fortney, J. J., Sudarsky, D., Hubeny, I., Cooper, C. S., Hubbard, W. B., Burrows, A., \& Lunine, J. I. 2003, \apj, 589, 615
\bibitem[Fortney et al.(2005)]{fortney et al 2005} Fortney, J. J., Marley, M. S., Lodders, K., Saumon, D. \& Freedman, R. 2005, \apj, 627, L69
\bibitem[Fortney et al.(2006)]{fortney et al 2006} Fortney, J. J., Saumon, D., Marley, M. S., Lodders, K., \& Freedman, R. S. 2006, \apj, 642, 495
\bibitem[Gladstone et al.(1996)]{gladstone et al 1996} Gladstone, G.~R., Allen, M., \& Yung, Y.~L.\ 1996, Icarus, 119, 1
\bibitem[Griffith \& Yelle(1999)]{griffith and yelle 1999} Griffith, C.~A., \& Yelle, R.~V.\ 1999, \apjl, 519, L85
\bibitem[Gurvich et al.(1989-1994)]{gurvich et al 1989-1994} Gurvich, L. V., Veyts, I. V., \& Alcock, C. B. 1989-1994, {Thermodynamic Properties of Individual Substances}, 4th ed., 3 vols. (New York: Hemisphere Publishing)
\bibitem[Gustafsson et al.(1999)]{gustafsson et al 1999} Gustafsson, B., Karlsson, T., Olsson, E., Edvardsson, B., \& Ryde, N.\ 1999, \aap, 342, 426
\bibitem[Huang et al.(2005)]{huang et al 2005} Huang, C., Zhao, G., Zhang, H. W., \& Chen, Y. Q.\ 2005, \mnras, 363, 71
\bibitem[Husain \& Marshall(1986)]{husain and marshall 1986} Husain, D., \& Marshall, P. 1986, Int. J. Chem. Kin., 18, 83
\bibitem[Iro et al.(2005)]{iro et al 2005} Iro, N., B\'{e}zard, B., \& Guillot, T. 2005, \aap, 436, 719
\bibitem[Kirkpatrick et al.(1999)]{kirkpatrick et al 1999} Kirkpatrick, J. D., et al. 1999, \apj, 519, 802
\bibitem[Kurbanov \& Mamedov(1995)]{kurbanov and mamedov 1995} Kurbanov, M. A., \& Mamedov, Kh. F. 1995, Kinet. Catal. 36, 455
\bibitem[Larson et al.(1984)]{larson et al 1984} Larson, H.~P., Bjoraker, G. L., Davis, D. S., \& Hofmann, R.\ 1984, \icarus, 60, 621
\bibitem[Liang et al.(2003)]{liang et al 2003} Liang, M. C., Parkinson, C. D., Lee, A. Y. T., Yung, Y. L., \& Seager, S. 2003, \apj, 596, L247
\bibitem[Liebert et al.(2000)]{liebert et al 2000} Liebert, J., Reid, I. N., Burrows, A., Burgasser, A. J., Kirkpatrick, J. D., \& Gizis, J. E. 2000, \apj, 533, L155
\bibitem[Lewis(1969a)]{lewis 1969a} Lewis, J. S. 1969, \icarus, 10, 393
\bibitem[Lewis(1969b)]{lewis 1969b} Lewis, J. S.\ 1969, Icarus, 10, 365
\bibitem[Lodders(1999a)]{lodders 1999a} Lodders, K. 1999a, \apj, 519, 793
\bibitem[Lodders(1999b)]{lodders 1999b} Lodders, K. 1999b, {J. Phys. Chem. Ref. Data}, 28, 1705
\bibitem[Lodders(2002)]{lodders 2002} Lodders, K. 2002, \apj, 577, 974
\bibitem[LF02()Lodders \& Fegley 2002]{lodders and fegley 2002} Lodders, K., \& Fegley, B., Jr. 2002, \icarus, 155, 393; LF02
\bibitem[Lodders(2003)]{lodders 2003} Lodders, K. 2003, \apj, 591, 1220
\bibitem[Lodders(2004a)]{lodders 2004a} Lodders, K. 2004a, {J. Phys. Chem Ref. Data}, 33, 357
\bibitem[Lodders(2004b)]{lodders 2004b} Lodders, K. 2004b, \apj, 611, 587
\bibitem[Lodders \& Fegley(2006)]{lodders and fegley 2006} Lodders, K., \& Fegley, B., Jr. 2006, in Astrophysics Update 2, ed. J. W. Mason (Springer-Praxis)
\bibitem[Marley et al.(1996)]{marley et al 1996} Marley, M. S., Saumon, D., Guillot, T., Freedman, R. S., Hubbard, W. B., Burrows, A., \& Lunine, J. I. 1996, {Science}, 272, 191
\bibitem[McLean et al.(2003)]{mclean et al 2003} McLean, I. S., McGovern, M. R., Burgasser, A. J., Kirkpatrick, J. D., Prato, L., \& Kim, S. S. 2003, \apj, 596, 561
\bibitem[Niemann et al.(1998)]{niemann et al 1998} Niemann H. B., et al. 1998, \jgr, 103, 22831
\bibitem[Noll et al.(1995)]{noll et al 1995} Noll, K. S., et al. 1995, Science, 267, 1307
\bibitem[Noll \& Marley(1997)]{noll and marley 1997} Noll, K.~S., \& Marley, M.~S.\ 1997, ASP Conf.~Ser.~119: Planets Beyond the Solar System and the Next Generation of Space Missions, 119, 115
\bibitem[Prinn \& Owen(1976)]{prinn and owen 1976} Prinn, R.~G., \& Owen, T.\ 1976, in Jupiter, ed. T. Gehrels (Tuscon: Univ. of Arizona Press), 319
\bibitem[Prinn \& Olaguer(1981)]{prinn and olaguer 1981} Prinn, R.~G., \& Olaguer, E.~P.\ 1981, \jgr, 86, 9895
\bibitem[Prinn et al.(1984)]{prinn et al 1984} Prinn, R. G., Larson, H. P., Caldwell, J. J., \& Gautier, D.\ 1984, in Saturn, ed. T. Gehrels \& M. S. Matthews (Tuscon: Univ. of Arizona Press), 88
\bibitem[Robie \& Hemingway(1995)]{robie and hemingway 1995} Robie, R. A., \& Hemingway, B. S. 1995, US Geological Survey Bull. 2131
\bibitem[Samland(1998)]{samland 1998} Samland, M.\ 1998, \apj, 496, 155
\bibitem[Saumon et al.(2000)]{saumon et al 2000} Saumon, D., Geballe, T. R., Leggett, S. K., Marley, M. S., Freedman, R. S., Lodders, K., Fegley, B., Jr., \& Sengupta, S. K. 2000, \apj, 541, 374
\bibitem[Saumon et al.(2003)]{saumon et al 2003} Saumon, D., Marley, M. S., \& Lodders, K. 2003, arXiv:astro-ph/0310805
\bibitem[Smith et al.(2001)]{smith et al 2001} Smith, V.~V., Cunha, K., \& King, J.~R.\ 2001, \aj, 122, 370
\bibitem[Timmes et al.(1995)]{timmes et al 1995} Timmes, F. X., Woosley, S. E., \& Weaver, T. A.\ 1995, \apjs, 98, 617
\bibitem[Tsuji(1973)]{tsuji 1973} Tsuji, T. 1973, \aap, 23, 411
\bibitem[Tsuji et al.(1996)]{tsuji et al 1996} Tsuji, T., Ohnaka, K., \& Aoki, W. 1996, \aap, 305, L1
\bibitem[Twarowski(1995)]{twarowski 1995} Twarowski, A. 1995, Combust. Flame, 102, 41
\bibitem[Visscher \& Fegley(2005)]{visscher and fegley 2005} Visscher, C., \& Fegley, B, Jr. 2005,  \apj, 623, 1221
\bibitem[West et al.(2004)]{west et al 2004} West, R.~A., Baines, K.~H., Friedson, A.~J., Banfield, D., Ragent, B., \& Taylor, F.~W.\ 2004, in Jupiter, ed. F. Bagenal, T.~E. Dowling, \& W.~B. McKinnon (Cambridge: Cambridge Univ. Press), 79
\bibitem[Woiki \& Roth(1994)]{woiki and roth 1994} Woiki, D., \& Roth, P. 1994, J. Phys. Chem., 98, 12958
\bibitem[Wong et al.(2004)]{wong et al 2004} Wong, M. H., Mahaffy, P. R., Atreya, S. K., Niemann, H. B., \& Owen, T. C. 2004, \icarus, 171, 153
\bibitem[Yamamura et al.(2000)]{yamamura et al 2000} Yamamura I., Kawaguchi, K., \& Ridgeway, S. T. 2000, \apj, 528, L33
\bibitem[Yelle(1999)]{yelle 1999} Yelle, R. 1999, in From Giant Planets to Cool Stars, ASP Conference Series, Vol. 212, ed. C. A. Griffith \& M. S. Marley (San Francisco: ASP), 267
\bibitem[Yoshimura et al.(1992)]{yoshimura et al 1992} Yoshimura, M, Koshi, M., Matsui, H., Kamiya, K., \& Umeyama, H. 1992, Chem. Phys. Lett., 189, 199
\bibitem[Zahnle et al.(1995)]{zahnle et al 1995} Zahnle, K., Mac Low, M.-M., Lodders, K., \& Fegley, B.\ 1995, \grl, 22, 1593
\end{thebibliography}
\end{document}